\documentclass[a4paper]{article}

\usepackage[english]{babel}
\usepackage[utf8x]{inputenc}
\usepackage[T1]{fontenc}
\usepackage{ upgreek }
\usepackage{gensymb} 
\makeatletter
\newcommand{\ssymbol}[1]{^{\@fnsymbol{#1}}}
\makeatother

\usepackage{amsmath}
\usepackage{braket}
\usepackage{graphicx}
\usepackage[colorinlistoftodos]{todonotes}
\usepackage[colorlinks=true, allcolors=blue]{hyperref}

\usepackage[a4paper,top=3cm,bottom=2cm,left=3cm,right=3cm,marginparwidth=1.75cm]{geometry}

\usepackage{authblk}
\title{Quantum oscillations in the black hole horizon}
\author[$\ssymbol{1}$]{C. Corda}
\author[$\ssymbol{2}$]{F. Feleppa}
\author[$\ssymbol{3}$]{F. Tamburini}
\author[$\ssymbol{1}$,$\ssymbol{4}$]{I. Licata}
\affil[$\ssymbol{1}$]{International Institute for Applicable Mathematics and Information Sciences (IIAMIS), B.M. Birla Science Centre, Adarsh Nagar, Hyderabad -- 500 463, India and Istituto Livi, Via Antonio Marini, 9, 59100 Prato (Italy)}
\affil[$\ssymbol{2}$]{Institute for Theoretical Physics, Utrecht University, Princetonplein 5, 3584 CC Utrecht, The Netherlands}
\affil[$\ssymbol{3}$]{ZKM -- Zentrum f\"ur Kunst und Medientechnologie, Lorentzstr. 19, D-76135, Karlsruhe, Germany}
\affil[$\ssymbol{4}$]{Institute for Scientific Methodology (ISEM), Palermo, Italy, School of Advanced International Studies on Theoretical and Nonlinear Methodologies of Physics, Bari, I-70124, Italy}

\begin{document}
\maketitle

\begin{abstract}
By applying Rosen's quantization approach to the historical Oppenheimer
and Snyder gravitational collapse and by setting the constraints for
the formation of the Schwarzschild black hole (SBH), in a previous
paper \cite{key-5} two of the Authors (CC and FF) found the gravitational potential, the Schr\"odinger
equation, the solution for the energy levels, the area quantum and
the quantum representation of the ground state at the Planck scale
of the SBH. Such results are consistent with previous ones in the
literature. It was also shown that the traditional
classical singularity in the core of the SBH is replaced by a quantum oscillator
describing a non-singular two-particle system where the two components, 
named the ``nucleus'' and the ``electron'', strongly interact with each other through a quantum
gravitational interaction. In agreement with the de Broglie hypothesis, the ``electron'' is interpreted in terms of the quantum oscillations of the BH horizon. In other words, the SBH should be the gravitational analogous of the hydrogen atom. In this paper, it is shown that these results allow us to compute the SBH entropy as a function of the BH principal quantum number in terms of Bekenstein-Hawking entropy and three sub-leading
corrections. In addition, the coefficient of the formula of Bekenstein-Hawking entropy is reduced to
a quarter of the traditional value. Then, it is shown that, by performing a correct rescaling of the energy levels, the semi-classical Bohr-like approach to BH quantum physics, previously developed by one of the Authors (CC), is consistent with the obtained results for large values of the BH principal quantum
number. After this, Hawking radiation will be analysed by discussing
its connection with the BH quantum structure. Finally, it is shown
that the time evolution of the above mentioned system solves the BH information paradox.

\end{abstract}

\section{Introduction}
Black holes are at all effects theoretical and conceptual laboratories where
one discusses, tests and tries to understand the fundamental problems
and potential contradictions that arise in the various attempts made
to unify Einstein's general theory of relativity with quantum mechanics.
In a previous paper \cite{key-5} two of the Authors (CC and FF) suggested a new model
of quantum BH based on a mathematical analogy to that of the hydrogen
atom obtained by using the same quantization approach proposed by the historical collaborator of Einstein, N. Rosen \cite{key-1}. It is well-known that the canonical quantization of general relativity leads to the Wheeler-DeWitt equation introducing the so-called Superspace, an infinite-dimensional space of all possible 3-metrics; Rosen, instead, preferred to start his work from the classical cosmological equations using a simplified quantization scheme. He wanted to reduce, at least formally, the cosmological Einstein-Friedman equations of general relativity to a quantum mechanical system; if this issue holds, the Friedman equations can be recast as a Schr\"odinger equation and the cosmological solutions can be read as eigensolutions of such a \textquotedblleft cosmological” Schr\"odinger equation. In this way Rosen found that, in the case of a Universe filled with pressureless matter, the equation is like that for the $s$ states of a hydrogen-like atom \cite{key-1}. It is important to recall that that attempts at quantising the FLRW universe date back at least to DeWitt's first of his famous 1967 papers \cite{key-75}, where the interesting observation is made that one needs a lot of particles to ensure the semiclassical behaviour.
\\
Actually, Rosen's approach can be also applied to the historical Oppenheimer and Snyder 
gravitational collapse \cite{key-2}, that is the simple case of a pressureless \textquotedblleft \emph{star of dust}\textquotedblright{}. By applying the constraints for the formation of the
SBH, in \cite{key-5} it has been found the gravitational potential,
the Schr\"odinger equation and the solution for the energy levels of
the black hole. The energy spectrum that has been found is consistent with the
one which was found by Bekenstein in 1974 \cite{key-11} and
with other ones in the literature \cite{key-12,key-13,key-14,key-15,key-16}. The black hole area
quantization has been also achieved by finding a result similar to
the one obtained by Bekenstein but with a different coefficient.
A quantum representation of the SBH ground state at the Planck scale has been obtained too. Finally,
it has been shown that the traditional classical singularity in the
core of the SBH is replaced, in a full quantum treatment,
by a two-particle system where the two components strongly interact
with each other through a quantum gravitational interaction.
This system seems to be non-singular from the quantum
point of view and it is analogous to the hydrogen atom because it consists
of a ``nucleus'' and an ``electron''. In agreement with the de Broglie hypothesis \cite{de Broglie}, the ``electron'' is interpreted in terms of the quantum oscillations of the BH horizon.
Let us remark that the study of the collapse of a simple set of dust particles can be a first
and valid approach to face the problem as it presents the advantage of finding and putting in evidence some fundamental properties that can be found also in other models characterized by, e.g.,
more articulated descriptions in terms of quantum fields that require a more complicated interpretation. This method, even if simplified, has a good validity as, up to now, there is not a unique
and valid theory of quantum gravity that suggests a specific model to use. On the other hand, one must also stress that a very complex phenomenon such as the gravitational collapse of a compact body is here treated in the highly idealised Oppenheimer and Snyder model, where only one degree of freedom survives and is quantised. This makes the emerging picture of quantum BHs oversimplified as well and really consistent only for Schwarzschild BHs, which are the final result of the Oppenheimer and Snyder gravitational collapse. Moreover, matter in the Oppenheimer and Snyder model is dust, and this is consistent with the hydrogen-like potential energy which will be found in the next Section, if one takes the Newtonian form. Then, that BHs then look like hydrogen atoms should be no surprise given the premises. For the sake of completeness, one recalls that, concerning the Oppenheimer and Snyder model,  some progress with respect to \cite{key-5} has been realized in the recent attempt \cite{key-74}.  
\\
This paper is organized as it follows. In Section II, Rosen's approach,
which has been discussed in \cite{key-5}, is reviewed by
also adding some interesting new insights; in Section III, it is shown
that the same results can be obtained also through a path integral approach; in Section IV, the SBH entropy is calculated in terms of Bekenstein-Hawking entropy and three sub-leading corrections.
It is also shown that the coefficient of the formula of Bekenstein-Hawking
entropy is reduced to one half of its traditional value;
in Section 5, it is shown that, by performing a rescaling
of the energy levels, the semi-classical Bohr-like approach to BH
quantum physics, previously developed by one of the Authors (CC) in
\cite{key-8,key-9} and reviewed in \cite{key-21}, is consistent
with the obtained results for large values of the BH principal quantum
number; in Section VI, Hawking radiation will be analysed by discussing
its connection with the BH quantum structure; in Section VII, we analyze the time evolution of the ``gravitational
hydrogen atom'' (GHA), and discuss how the resolution of the BH information paradox can be achieved; finally, Section
VIII is devoted to the conclusion remarks.
\section{A review of Rosen's approach}
Classically, the gravitational collapse in the simple case of a pressureless
\textquotedblleft star of dust\textquotedblright{} with uniform density
is well known \cite{key-4}. From the historical point of view, it
was originally analysed in the famous paper of Oppenheimer and Snyder
\cite{key-2}. For the interior of the collapsing star, one indeed uses the well-known Friedmann-Lemaitre-Robertson-Walker (FLRW) line-element with comoving hyper-spherical coordinates $\chi,$
$\theta,$ $\varphi$ \cite{key-4}. Thus, one writes down \cite{key-4}
(hereafter Planck units will be used, i.e. $G=c=k_{B}=\hbar=\frac{1}{4\pi\epsilon_{0}}=1$)

\begin{equation}
ds^{2}=d\tau^{2}+a(\tau)(-d\chi^{2}-\sin^{2}\chi(d\theta^{2}+\sin^{2}\theta d\varphi^{2}),\label{eq: metrica conformemente piatta}
\end{equation}
where the origin of coordinates is set at the centre of the star,
and $a(\tau)$ is the scale factor given by the familiar cycloidal
relation \cite{key-4}
\begin{equation}
\begin{array}{c}
a=\frac{1}{2}a_{m}\left(1+\cos\eta\right),\\
\\
\tau=\frac{1}{2}a_{m}\left(\eta+\sin\eta\right),
\end{array}\label{eq: cycloidal relation}
\end{equation}
while the density is given by \cite{key-4}
\begin{equation}
\rho=\left(\frac{3a_{m}}{8\pi}\right)a^{-3}=\left(\frac{3}{8\pi a_{m}^{2}}\right)\left[\frac{1}{2}\left(1+\cos\eta\right)\right]^{-3}.\label{eq: density}
\end{equation}
Setting $\sin^{2}\chi$ one chooses the case of positive curvature,
which corresponds to a gas sphere whose dynamics begins at rest with
a finite radius, and, in turn, it is the only one of interest \cite{key-4}.
Thus, the choice $k=1$ is made for dynamical reasons (the initial
rate of change of density is null, that means ``momentum of maximum
expansion'' \cite{key-4}), but the dynamics also depends on the
field equations. 

In order to discuss the simplest model of a \emph{``star of dust''}, that is, the case of zero pressure, one sets the stress-energy tensor as \cite{key-4}
\begin{equation}
T=\rho u\otimes u,\label{eq: stress energy}
\end{equation}
where $\rho$ is the density of the collapsing star and $u$ the four-vector
velocity of the matter. On the other hand, the external geometry is
given by the Schwartzschild line-element \cite{key-4}
\begin{equation}
ds^{2}=\left(1-\frac{2M}{r}\right)dt^{2}-r^{2}\left(\sin^{2}\theta d\varphi^{2}+d\theta^{2}\right)-\frac{dr^{2}}{1-\frac{2M}{r}},\label{eq: Hilbert}
\end{equation}
where $M$ is the total mass of the collapsing star which is constant
during the collapse \cite{key-4}. As there are no pressure gradients,
which can deflect the particles motion, the particles on the surface
of any ball of dust move along radial geodesics in the exterior Schwarzschild
space-time \cite{key-4}. Considering a ball which begins at rest
with finite radius (in terms of the Schwarzschild radial coordinate)
$r=r_{i}$ at the (Schwarzschild) time $t=0,$ the geodesics motion
of its surface is given by the following equations \cite{key-4}:
\begin{align}
r&=\frac{1}{2}r_{i}\left(1+\cos\eta\right),\label{eq: geodesics surface radius}\\
t&=2M\ln\left[\frac{\sqrt{\frac{r_{i}}{2M}-1}+\tan\left(\frac{\eta}{2}\right)}{\sqrt{\frac{r_{i}}{2M}-1}-\tan\left(\frac{\eta}{2}\right)}\right]\\
&\hspace{0.4cm}+2M\sqrt{\frac{r_{i}}{2M}-1}\left[\eta+\left(\frac{r_{i}}{4M}\right)\left(\eta+\sin\eta\right)\right].
\end{align}\label{eq: geodesic surface time}
The proper time measured by a clock put on the surface of the collapsing
star is \cite{key-4}
\begin{equation}
\tau=\sqrt{\frac{r_{i}^{3}}{8M}}\left(\eta+\sin\eta\right).\label{eq: tau}
\end{equation}
The collapse begins for $r=r_{i},$ $\eta=\tau=t=0,$ and terminates
at the singularity $r=0,$ $\eta=\pi$ after a duration of proper
time measured by the falling particles \cite{key-4}
\begin{equation}
\Delta\tau=\pi\sqrt{\frac{r_{i}^{3}}{8M}},\label{eq: delta tau}
\end{equation}
which coincidentally corresponds, as it is well known, to the interval
of Newtonian time for free-fall collapse in Newtonian theory. Differently
from the cosmological case, where the solution is homogeneous and
isotropic everywhere, here the internal homogeneity and isotropy of
the FLRW line-element are broken at the star's surface, that is, a
some radius $\chi=\chi_{0}$ \cite{key-4}. At that surface, which
is a 3-dimensional world tube enclosing the star's fluid \cite{key-4},
the interior FLRW geometry must match smoothly the exterior Schwarzschild
geometry \cite{key-4}. One considers a range of $\chi$ given by
$0\leq\chi\leq\chi_{0}$, with $\chi_{0}<\frac{\pi}{2}$ during the
collapse \cite{key-4}. For the pressureless case the match is possible
\cite{key-4}. The external Schwarzschild solution predicts indeed
a cycloidal relation for the star's circumference \cite{key-4}
\begin{equation}
\begin{array}{c}
C=2\pi r=2\pi\left[\frac{1}{2}r_{i}\left(1+\cos\eta\right)\right],\\
\\
\tau=\sqrt{\frac{r_{i}^{3}}{8M}}\left(\eta+\sin\eta\right).
\end{array}\label{eq: Circonferenza esterna}
\end{equation}
The interior FLRW predicts a similar cycloidal relation \cite{key-4}
\begin{equation}
\begin{array}{c}
C=2\pi r=2\pi a\sin\chi_{0}=\pi\sin\chi_{0}a_{m}\left(1+\cos\eta\right),\\
\\
\tau=\frac{1}{2}a_{m}\left(\eta+\sin\eta\right).
\end{array}\label{eq: circonferenza esterna}
\end{equation}
Therefore, the two predictions agree perfectly for all time if and
only if \cite{key-4} 

\begin{equation}
\begin{array}{c}
r_{i}=a_{m}\sin\chi_{0},\\
\\
M=\frac{1}{2}a_{m}\sin^{3}\chi_{0,}
\end{array}\label{eq: matching}
\end{equation}
where $r_{i}$ and $a_{m}$ are the values of the Schwarzschild radial
coordinate in Eq. (\ref{eq: Hilbert}) and of the scale factor in
Eq. (\ref{eq: metrica conformemente piatta}) at the beginning of
the collapse, respectively. Thus, Eqs. (\ref{eq: matching}) represent
the requested match, while the Schwarzschild radial coordinate, in
the case of the matching between the internal and external geometries,
is \cite{key-4}

\begin{equation}
r=a\sin\chi_{0}.\label{eq: mach}
\end{equation}

One underlines that the Oppenheimer and Snyder gravitational collapse
is not physical and can be considered as a toy model. But here the key
point is that, as it is well known from the historical paper of Oppenheimer
and Snyder \cite{key-2} and by various subsequent works, the final
state of this simplified gravitational collapse is the SBH, which,
instead, has a fundamental role in quantum gravity. It is indeed a
general conviction, arising from an idea of Bekenstein \cite{key-6},
that, in the search of a quantum gravity theory, the SBH should play
a role similar to the hydrogen atom in quantum mechanics. It should
be a ``theoretical laboratory'' where one discusses and tries to
understand conceptual problems and potential contradictions in the
attempts to unify Einstein's general relativity with quantum mechanics.
Thus, despite non-physical, the Oppenheimer and Snyder gravitational
collapse must be here considered as a tool which permits to understand
a fundamental physical system, that is the SBH. In fact, it will be
shown that, by setting the constraints for the formation of the SBH
in the quantized Oppenheimer and Snyder gravitational collapse, one
arrives to quantize the SBH, and this will be a remarkable, important
result in the quantum gravity's search. In fact, the Oppenheimer and Snyder gravitational collapse and/or some of its modifications have currently a renovated interest among researchers in gravitation, see for example \cite{key-67,key-68,key-69,key-70,key-71,key-72,key-73} and references within.

By rewriting the FLRW line-element (\ref{eq: metrica conformemente piatta})
in spherical coordinates and comoving time as
\begin{equation}
ds^{2}=d\tau^{2}-a^{2}(\tau)\left(\frac{dr^{2}}{1-r^{2}}+r^{2}d\theta^{2}+r^{2}\sin^{2}\theta d\varphi^{2}\right),\label{eq: FLRW}
\end{equation}
and by using the Einstein field equation
\begin{equation}
G_{\mu\nu}=-8\pi T_{\mu\nu},\label{eq: Einstein field equation}
\end{equation}
one gets the dynamical equations for the Oppenheimer and Snyder gravitational
collapse as \cite{key-5,key-1}
\begin{equation}
\begin{array}{c}
\dot{a}^{2}=\frac{8}{3}\pi a^{2}\rho-1, \hspace{0.2cm}
\ddot{a}=-\frac{4}{3}\pi a\rho,
\end{array}\label{eq: evoluzione}
\end{equation}
with $\dot{a}=\frac{da}{d\tau}$. Following \cite{key-5,key-1}, for
consistency, one gets 
\begin{equation}
\frac{d\rho}{da}=-\frac{3\rho}{a},\label{eq: consistenza}
\end{equation}
which, when integrated, gives 
\begin{equation}
\rho=\frac{C}{a^{3}}.\label{eq: densit=0000E0}
\end{equation}
The constant $C$ is determined by the initial conditions and it turns out to be
\begin{equation}
C=\frac{3a_{0}}{8\pi}.\label{eq: C}
\end{equation}
Thus, one rewrites Eq. (\ref{eq: densit=0000E0}) as 
\begin{equation}
\rho=\frac{3a_{0}}{8\pi a^{3}}.\label{eq: densit=0000E0 2}
\end{equation}
The first of (\ref{eq: evoluzione}) can be written as
\begin{equation}
\frac{1}{2}M\dot{a}^{2}-\frac{4}{3}\pi Ma^{2}\rho=-\frac{M}{2},\label{eq: energy equation for a particle}
\end{equation}
which seems like the energy equation for for a particle in one-dimensional
motion having coordinate $a$:
\begin{equation}
E=T+V,\label{eq: energia totale}
\end{equation}
where 
\begin{equation}
T=\frac{M\dot{a}^{2}}{2}\label{eq: energia cinetica}
\end{equation}
and 
\begin{equation}
V(a)=-\frac{4}{3}\pi Ma^{2}\rho,\label{eq: energia potenziale}
\end{equation}
are the kinetic and potential energy, respectively. Then, the total energy is 
\begin{equation}
E=-\frac{M}{2}\label{eq: energia totale 2}
\end{equation}
From the second of Eqs. (\ref{eq: evoluzione}) one gets the equation of motion of this particle, that is
\begin{equation}
M\ddot{a}=-\frac{4}{3}M\pi a\rho.\label{eq: equation of motion}
\end{equation}
The momentum of the particle is 
\begin{equation}
P=M\dot{a},\label{eq: momentum}
\end{equation}
with an associated Hamiltonian 
\begin{equation}
H=\frac{P^{2}}{2M}+V.\label{eq: Hamiltonian}
\end{equation}
Till now the discussion has been classical. In order to start the
quantum analysis, one needs to define a wave-function as \cite{key-5,key-1}
\begin{equation}
\Psi\equiv\Psi\left(a,\tau\right).\label{eq: wave-function}
\end{equation}
Hence, in correspondence of the classical equation (\ref{eq: Hamiltonian}),
one finds the traditional Schr\"odinger equation: 
\begin{equation}
i\frac{\partial\Psi}{\partial\tau}=-\frac{1}{2M}\frac{\partial^{2}\Psi}{\partial a^{2}}+V\Psi.\label{eq: Schrodinger equation}
\end{equation}
For a stationary state with energy $E$ one gets 
\begin{equation}
\Psi=\Psi\left(a\right)\exp\left(-iE\tau\right),\label{eq: separazione}
\end{equation}
and Eq. (\ref{eq: wave-function}) becomes
\begin{equation}
-\frac{1}{2M}\frac{\partial^{2}\Psi}{\partial a^{2}}+V\Psi=E\Psi.\label{eq: Schrodinger equation 2}
\end{equation}
Inserting Eq. (\ref{eq: densit=0000E0 2}) into Eq. (\ref{eq: energia potenziale})
one obtains 
\begin{equation}
V(a)=-\frac{Ma_{0}}{2a}.\label{eq: energia potenziale 2}
\end{equation}
Setting 
\begin{equation}
\Psi=aX,\label{eq: X}
\end{equation}
Eq. (\ref{eq: Schrodinger equation 2}) becomes 
\begin{equation}
-\frac{1}{2M}\left(\frac{\partial^{2}X}{\partial a^{2}}+\frac{2}{a}\frac{\partial X}{\partial a}\right)+VX=EX.\label{eq: Schrodinger equation 3}
\end{equation}
With $V$ given by Eq. (\ref{eq: energia potenziale 2}), Eq. (\ref{eq: Schrodinger equation 3}),
is analogous to the Schr\"odinger equation in polar coordinates for
the $s$ states ($l=0$) of a hydrogen-like atom \cite{key-5,key-1,key-7}
in which the squared electron charge $e^{2}$ is replaced by $\frac{Ma_{0}}{2}$.
Thus, for the bound states ($E<0$) the energy spectrum is 
\begin{equation}
E_{n}=-\frac{a_{0}^{2}M^{3}}{8n^{2}},\label{eq: spettro energia}
\end{equation}
where $n$ is the principal quantum number. Following again \cite{key-5,key-1},
one inserts Eq. (\ref{eq: energia totale 2}) into Eq. (\ref{eq: spettro energia}),
obtaining the mass spectrum of the gravitational collapse: 
\begin{equation}
M_{n}=\frac{a_{0}^{2}M_{n}^{3}}{4n^{2}}\Rightarrow M_{n}=\frac{2n}{a_{0}}.\label{eq: spettro massa}
\end{equation}
On the other hand, by using Eq. (\ref{eq: energia totale 2}), one
finds the energy levels of the collapsing star as 
\begin{equation}
E_{n}=-\frac{n}{a_{0}}.\label{eq: energy levels}
\end{equation}
Eq. (\ref{eq: spettro massa}) represents the spectrum of the total
mass of the collapsing star, while Eq. (\ref{eq: energy levels})
represents the energy spectrum of the the collapsing
star where the gravitational energy, which is given by Eq. (\ref{eq: energia potenziale 2}),
is included. The total energy of a quantum system with bound states
is indeed negative. Concerning the gravitational energy, one recalls
that, in general, the equivalence principle forbids its localization
in general relativity, with the sole exception of a spherical star
\cite{key-4}, which is exactly the case analysed in this paper. In
this case the gravitational energy is indeed localized not by mathematical
conventions, but by the circumstance that transfer of energy is detectable
by local measures \cite{key-4}. Thus, one can surely consider Eq.
(\ref{eq: energia potenziale 2}) as the gravitational potential energy
of the collapsing star.

Now, let us consider the case of a completely collapsed star, i.e.,
a SBH;  this means $\chi_{0}=\frac{\pi}{2}$, $r=a$ and $r_{i}=a_{0}=2M=r_{g},$
in Eqs. (\ref{eq: matching}) and (\ref{eq: mach}). Therefore, Eqs. (\ref{eq: energia potenziale 2}) and (\ref{eq: X}) become 
\begin{align}
V(r)&=-\frac{M^{2}}{r},\label{eq: 30}\\
\Psi&=rX\label{eq: 31},
\end{align}
Eq. (\ref{eq: Schrodinger equation 3}) become
\begin{equation}
-\frac{1}{2M}\left(\frac{\partial^{2}X}{\partial r^{2}}+\frac{2}{r}\frac{\partial X}{\partial r}\right)+VX=EX,\label{eq: Schrodinger equation BH}
\end{equation}
with
\begin{align}
E_{n}&=-\frac{r_{g}^{2}M^{3}}{8n^{2}},\label{eq: 33}\\
M_{n}&=\sqrt{n},\label{eq: 34}\\
E_{n}&=-\sqrt{\frac{n}{4}}.\label{eq: 35}
\end{align}
Eqs. (\ref{eq: 30}), (\ref{eq: Schrodinger equation BH}), (\ref{eq: 34}) and (\ref{eq: 35})
should be the exact gravitational potential energy\emph{,} Schr\"odinger
equation, mass spectrum and energy spectrum for the SBH interpreted
as GHA, respectively. Actually, a further
final correction is needed. To better clarify this point, one compares our Eq. (\ref{eq: 30}) with the analogous potential energy
of an hydrogen atom, which is \cite{key-7} 
\begin{equation}
V(r)=-\frac{e^{2}}{r}.\label{eq: energia potenziale atomo idrogeno}
\end{equation}
Eqs. (\ref{eq: 30}) and (\ref{eq: energia potenziale atomo idrogeno})
are formally identical, but there is an important difference. In (\ref{eq: energia potenziale atomo idrogeno}), the electron's
charge is constant for all the energy levels of the hydrogen atom.
Instead, in the case of Eq. (\ref{eq: 30}), based
on the emissions of Hawking quanta or on the absorptions of external
particles, the BH mass changes during the jumps from an energy level
to another. In fact, such a BH mass decreases for emissions and increases
for absorptions. Thus, one must also consider this BH dynamical behavior.
A good way to take into account the BH dynamical behavior is by introducing
the \emph{BH effective state}. This consists
in introducing some \emph{effective quantities}. Considering the initial
BH mass before a BH transition (an emission of a Hawking quantum or
an absorption of an external particle), $M$, and the final BH mass
after the transition, $M\pm\omega$, where $\omega$ is the mass-energy
of the particle involved in the transition (the sign $+$ or $-$ corresponds
to an absorption or to an emission, respectively), one introduces
the BH\emph{ effective mass and effective horizon }as \cite{key-5,key-8,key-9}
\begin{equation}
M_{E}\equiv M\pm\frac{\omega}{2},\mbox{ }r_{E}\equiv2M_{E},\label{eq: effective quantities absorption}
\end{equation}
respectively. The effective quantities (\ref{eq: effective quantities absorption})
represent average quantities. The variable \emph{$r_{E}$ }is indeed
the average of the initial and final horizons, while \emph{$M_{E}$
}is the average of the initial and final masses (see  \cite{key-5,key-8,key-9} for further details).
They represent the BH mass and horizon\emph{ during} the BH contraction
(expansion), i.e., \emph{during} the emission (absorption) of a particle.
Thus, on one hand, the introduction of the effective quantities (\ref{eq: effective quantities absorption})
in the BH dynamical equations is very intuitive. On the other hand, it is \emph{rigorously} justified through Hawking
periodicity argument \cite{key-10}. In order to take the BH dynamical behavior into due account,
one must replace the BH mass $M$ with the BH effective mass $M_{E}$
in Eqs. (\ref{eq: 30}), (\ref{eq: Schrodinger equation BH}),
(\ref{eq: 33}), and (\ref{eq: energia totale 2}),
obtaining \cite{key-5}
\begin{equation}
V(r)=-\frac{M_{E}^{2}}{r},\label{eq: energia potenziale BH effettiva}
\end{equation}
and
\begin{equation}
-\frac{1}{2M_{E}}\left(\frac{\partial^{2}X}{\partial r^{2}}+\frac{2}{r}\frac{\partial X}{\partial r}\right)+VX=EX,\label{eq: Schrodinger equation BH effettiva}
\end{equation}
with
\begin{align}
E_{n}&=-\frac{r_{E}^{2}M_{E}^{3}}{8n^{2}},\label{eq: 40}\\
E&=-\frac{M_{E}}{2}\label{eq: energia totale effettiva}.
\end{align}
Now, from the quantum point of view, one wants to obtain the energy
eigenvalues as being absorptions starting from the BH formation, that
is from the BH having null mass, where with ``the BH having null
mass'' one means the situation of the gravitational collapse before
the formation of the first event horizon. This implies that one must
replace $M\rightarrow0$ and $\omega\rightarrow M$ in Eqs. (\ref{eq: effective quantities absorption})
and must take the plus sign (absorptions) in the same equations. Thus,
one gets
\begin{equation}
M_{E}\equiv\frac{M}{2},\mbox{ }r_{E}\equiv2M_{E}=M.\label{eq: effective quantities absorption finali}
\end{equation}
Then, one inserts Eqs. (\ref{eq: energia totale effettiva}) and (\ref{eq: effective quantities absorption finali})
into Eq. (\ref{eq: 40}), obtaining the
BH mass spectrum as 
\begin{equation}
M_{n}=2\sqrt{n},\label{eq: spettro massa BH finale}
\end{equation}
and, by using Eq. (\ref{eq: energia totale effettiva}), one finds the
BH energy levels as 
\begin{equation}
E_{n}=-\sqrt{\frac{n}{4}}.\label{eq: BH energy levels finale.}
\end{equation}
In its absolute value, Eq. (\ref{eq: BH energy levels finale.}) is consistent with the BH
energy spectrum found by Bekenstein in 1974 \cite{key-11}. Bekenstein
indeed obtained $E_{n}\sim\sqrt{n}$ by using the Bohr-Sommerfeld
quantization condition because he argued that the SBH behaves as an
adiabatic invariant. In our opinion,
the Authors of previous literature did not consider that the BH energy
spectrum \emph{must have negative eigenvalues} because the GHA
 is a quantum system composed by bound states. The total BH
energy $E=-\frac{M}{4}$ \emph{is negative and different} from the
BH inert mass $M$. In fact, the quantization procedure in \cite{key-5}
splits the classical BH in a two-particle quantum system having a
gravitational negative energy given by Eq. (\ref{eq: energia potenziale BH effettiva}).
It is indeed well known that the bound states of physical systems
must have negative energies. 

Now, one recalls that in the SBH the \emph{horizon area} $A$ is related
to the mass by the relation $A=16\pi M^{2}.$ If one assumes that
a neighboring particle is captured by the BH causing a transition
from the state with $n$ to the state with $n+1$ (two neighboring
levels) by using (\ref{eq: spettro massa BH finale}),
one immediately gets the area quantum as 
\begin{equation}
\Delta A_{n\rightarrow n+1}=64\pi\left(n+1-n\right)=64\pi,\label{eq: area spectrum}
\end{equation}
which is similar to the original result of Bekenstein \cite{key-11},
but with a different coefficient. This is not surprising since
there is no general consensus on the area quantum. 
For the BH ground state ($n=1$), from Eq. (\ref{eq: spettro massa BH finale})
one gets
\begin{equation}
M_{1}=2\label{eq: massa minima}
\end{equation}
in Planck units. Therefore, in standard units, one gets $M_{1}=2m_{P},$
where $m_{P}$ is the Planck mass.
A total negative energy arising from Eq. (\ref{eq: BH energy levels finale.})
and a Schwarzschild radius are associated to the mass (\ref{eq: massa minima}):
\begin{align}\label{eq: energia minima}
E_{1}&=-\frac{1}{2},\\
r_{g1}&=4.
\end{align}
This is the state having minimum mass and minimum energy (the energy
of this state is minimum in absolute value; in its real value, being
negative, it is maximum). It represents the smallest possible BH.
In the case of Bohr's semi-classical model of hydrogen atom, the Bohr
radius, which represents the classical radius of the electron at the
ground state, is \cite{key-7} 
\begin{equation}
Bohr\:radius=\frac{1}{m_{e}e^{2}},\label{eq: Bohr radius}
\end{equation}
where $m_{e}$ is the electron mass. In order to obtain the correspondent
``Bohr radius'' for what we call the GHA, one needs
to replace both $m_{e}$ and $e$ in Eq. (\ref{eq: Bohr radius})
with the effective mass of the BH ground state, which is $\frac{M_{1}}{2}=1.$
Thus, now the ``Bohr radius'' becomes 
\begin{equation}
b_{1}=1,\label{eq: Bohr radius-1}
\end{equation}
which in standard units reads $b_{1}=l_{P},$ where $l_{P}$ is the Planck length. Thus, it has been found that the ``Bohr radius'' associated to the smallest possible BH is equal to the Planck length.
Following again \cite{key-1,key-5}, the wave-function associated to the
BH ground state is 
\begin{equation}
\Psi_{1}=2b_{1}^{-\frac{3}{2}}r\exp\left(-\frac{r}{b_{1}}\right)=2r\exp\left(-r\right),\label{eq: wave-function 1 BH}
\end{equation}
where $\Psi_{1}$ is normalized as
\begin{equation}
\int_{0}^{\infty}\Psi_{1}^{2}dr=1.\label{eq: normalizzazione BH}
\end{equation}
The size of this BH is of the order of 
\begin{equation}
\bar{r}_{1}=\int_{0}^{\infty}\Psi_{1}^{2}rdr=\frac{3}{2}b_{1}=\frac{3}{2}.\label{eq: size BH 1}
\end{equation}
It is also important to recall that in quantum mechanics the Bohr
radius (\ref{eq: Bohr radius-1}) gives the radius with the maximum
radial probability density instead of its expected radial distance
\cite{key-50}. The latter is indeed $1.5$ times the Bohr radius; this depends on the long tail of the radial wave function
and it is so given by Eq. (\ref{eq: size BH 1}). Thus,
remarkably, Eqs. (\ref{eq: Bohr radius-1}) and (\ref{eq: size BH 1})
are in complete agreement with the standard quantum approach to the
hydrogen atom.

Hence, an interesting quantum representation of the Schwarzschild
BH ground state at the Planck scale has been obtained. This Schwarzschild
BH ground state represents the BH minimum energy level which is compatible
with the generalized uncertainty principle (GUP) \cite{key-19}. The
GUP indeed prevents a BH from its total evaporation by stopping Hawking's
evaporation process in exactly the same way that the usual Heisenberg
uncertainty principle (HUP) prevents the hydrogen atom from total
collapse \cite{key-17}.

Following again Rosen \cite{key-1}, one can also easily find the
size, that is the "expected radial distance", of the quantum SBH excited
at the level $n$ as 
\begin{equation}
\bar{r}_{n}=\frac{3}{2}\sqrt{n}.\label{eq: size BH n}
\end{equation}

Let us clarify an important point. It is well known that
the Schwarzschild coordinates break down when the radius of the star
becomes the gravitational radius, which means that it is no longer
possible to match the $\chi_{0}$ surface with a space-like normal
to a constant $r$ surface which is null at the gravitational radius
and has a time-like normal for values of $r$ which are less than
the gravitational radius. Thus, we want to stress here that we never went beyond the gravitational radius. This
issue is clarified as follows. The passage from a classical to a quantum
analysis splits the classical BH in a two-particle system, the ``nucleus''
and the ``electron''. Here the key point is the physical meaning
of the second particle, the ``electron'' during the BH formation
and in the following BH dynamical evolution. From a quantum point
of view, the BH formation is interpreted as being the formation of
the initial pair composed by the ``nucleus'' and the ``electron''.
Thus, it represents an absorption from the BH having null mass. This
is not an istantaneous process, but, instead, it happens in a finite
lapse of time. Let us assume that, during a little, but finite, lapse
of comoving time (which is the time that we used in the above analysis),
say $\Delta\tau,$ the gravitational collapse generates a BH having
an initial mass, say $M_{n}=2\sqrt{n}.$ During $\Delta\tau$ the
BH is forming, which means that the two particles are evolving. When
the BH is completely formed the two particles are in a stationary
state, having an energy $E_{n}=-\frac{1}{2}\sqrt{n}.$ One notes that
in the two-particle quantum system, the two particles are equal and
can be mutually exchanged without varying the physical properties
of the system. This means that, when the system is in the stationary
state, one can assign to the ``electron'' the half of the total
energy of the two-particle quantum system, obtaining 
\begin{equation}
\frac{E_{n}}{2}=-\frac{\sqrt{n}}{4}=-\frac{n}{2M_{n}},\label{eq: energia elettrone}
\end{equation}
which, remarkably, corresponds exactly to a particle quantized with
antiperiodic boundary conditions on a circle of length
\begin{equation}
L=4\pi M_{n}=8\pi\sqrt{n}.\label{eq: lunghezza cerchio corretta 1}
\end{equation}
This means that the radius of the circle is $r_{g}=2M_{n},$ i.e.,
exactly the gravitational radius. Hence, when the system is in the
stationary state, that is, when the BH is completely formed, the distance
between the ``nucleus'' and the ``electron'' in terms of the Schwarzschild
radial coordinate is exactly the gravitational radius. 

Following the analogy with the hydrogen atom, one can evoke the de
Broglie hypothesis \cite{de Broglie} and consider the wave nature of
the BH ``electron''. This means that such particle does not orbit
the nucleus in the same way as a planet orbits the Sun, but instead
exists as standing wave. The correct analogy is that of a large and
often oddly shaped ``atmosphere'' (the BH
``electron''), distributed around a relatively tiny planet (the
BH ``nucleus''). The correct physical intepretation of such an ``atmosphere''
is nothing else than the BH horizon modes. In fact, the idea that
the radius of the event horizon undergoes quantum oscillations has
a longstanding history. Such horizon modes were introduced in a semi-classical
framework about 50 years ago \cite{Press} in terms of BH quasi-normal
modes (QNMs) which represent the BH back reaction to perturbations.
Both of the absorptions of external particles and the emissions of
Hawking quanta are BH perturbations and this allowed one of us,
CC, to develop the Bohr-like approach to BH quantum physics which
will be discussed and refined in Section 5 of this paper, starting
from an original idea of Hod \cite{key-38} which has been improved
by Maggiore \cite{key-39}. On one hand, the QNMs approach is a semi-classical
approach similar to the approach that Bohr developed in 1913 \cite{key-26,key-27}
concerning the structure of the hydrogen atom. On the other hand,
the importance of horizon modes in a quantum gravity framework
has been recently emphasized by considering them as being
described by the periodic motion of their particle-like analogue \cite{Spallucci},
in full accordance with the de Broglie hypothesis. The Authors found an energy spectrum which scales as which scales as $\sim\sqrt{n}$, which is consistent with the results obtained in this paper. The key point is 
that during both the processes of absorptions of external particles,
included the original BH formation, and of emissions of Hawking quanta,
the BH horizon is not fixed at a constant distance from the BH nucleus
\cite{key-12}. In fact, due to energy conservation, the BH contracts
during the emission of a particle and expands during an absorption
\cite{key-12}. Such quantum contractions/expansions are not ``one
shot processes'' \cite{key-12}. They generates oscillations of the
horizon instead \cite{key-12}. Thus, the ``Bohr radius'', Eq. (\ref{eq: Bohr radius-1}), and the ``expected radial distance'', Eq.  (\ref{eq: size BH n}), must be interpreted
as dynamical quantities characterizing the quantum state of the horizon
modes during a transition from a stationary state to another, while
the ``Bohr orbits'' of the stationary states are characterized by
Eq. (\ref{eq: lunghezza cerchio corretta 1}). Clearly, if the second
particle, that is the ``electron'', is interpreted in terms of
horizon oscillations, then the distance between the two particle is never less than the gravitational radius which oscillates. 

Another fundamental issue is the following. The quantum BH expressed
by the system of Eqs. (\ref{eq: energia potenziale BH effettiva}) - (\ref{eq: energia totale effettiva}) seems to be \emph{non-singular}.
It is indeed well known that, in the classical general relativistic
framework, in the internal geometry all time-like radial geodesics
of the collapsing star terminate after a lapse of finite proper time
in the termination point $r=0$ and it is impossible to extend the
internal space-time manifold beyond that termination point \cite{key-4}.
Thus, the point $r=0$ represents a singularity based on the
definition by Schmidt \cite{key-18}. But what happens in the quantum
framework in \cite{key-5} is completely different. The above analysis
has shown that the classical SBH 
is described in terms of a quantum harmonic oscillator that reflects
several analogies to that describing a two-particle
system where the two components strongly interact with each other
through a quantum gravitational interaction. 
The system that has been analysed so far is indeed formally 
equal to the well known system of two quantum particles having 
finite distance with the mutual attraction of the form $1/r$ \cite{key-7}. 
Similarly to the an hydrogen atom, this oscillator describes the
behaviour of a two particle system where one particle behaves as the ``nucleus''
and the other as an ``electron'' of a GHA. 
Of course, following this mathematical analogy, one has to figure out what
is the meaning of the word ``particle'' in a quantum framework.
Quantum particles remain in an uncertain, non-deterministic, smeared,
probabilistic wave-particle orbital state \cite{key-7}. 
Then, they cannot be localized in a particular ``termination point where all
time-like radial geodesics terminate'' \cite{key-5}. Such a localization
is also in contrast with the HUP, which says indeed that either
the location or the momentum of a quantum particle such as the BH
``electron'' can be known as precisely as desired, but as one of
these quantities is specified more precisely, the value of the other
becomes increasingly indeterminate \cite{key-5}. 
This is not simply a matter of observational difficulty, but 
this mathematical structure rather reflects a fundamental property
of nature. What this means is that within the tiny confines of the
``gravitational atom'', the so-called ``electron'' cannot really be regarded
as a ``point-like particle'' having a definite energy and location 
or an effective pair of particles as occurs in the real hydrogen atom \cite{key-5}. 
Thus, it is somewhat misleading to talk about the BH ``electron'' ``falling into''
the BH ``nucleus''. 
In other words, the Schwarzschild radial coordinate cannot become equal to
zero. The GUP makes this last statement even stronger. 
In fact, it can be written down in the general form as (see \cite{key-19} and the review \cite{key-78})
\begin{equation}
\Delta x\Delta p\geq\frac{1}{2}\left[1+\eta\left(\Delta p\right)^{2}+\ldots \right].\label{eq: GUP}
\end{equation}
Eq. (\ref{eq: GUP}) implies a non-zero lower
bound on the minimum value of the uncertainty on the particle's position which is of order of the Planck length \cite{key-19,key-78}. In other words, the GUP implies the existence of a minimal length $L$ in quantum gravity  also when Einstein's equations hold down to the Planck scales \cite{erepr}.
Following this hypothesis, instead of focusing on the energy--tensor quantity such as the momentum vector,  we obtain an equivalent indetermination relationship from an invariant scalar quantity, the proper energy $E$.
To obtain $E$ one integrates and averages Einstein's equations over a 3D space-like hypersurface, $\sigma$,
with unit normal vector $n\sim g^{-1/2}$. 

The proper energy is averaged over a proper volume $L^3$, 
and is obtained from the integral of the energy momentum tensor taken over a chosen proper volume 
element of a space-like 3D hypersurface, giving 
\begin{equation}
\bar E\sim \frac{g^2}{L}R_{(4)}= L \left( \Delta\left(\Delta g (g)^{-1}\right) + \left(\Delta g (g)^{-1}\right)^2\right).
\label{eqheisen}
\end{equation}
Rescaling Eq. (\ref{eqheisen}) down to the Planck scale $L_p$ and defining, in natural units, the light crossing time 
$\tau = L$ and the Planck Time $\tau_p$, 
one obtains the indetermination relationship valid down to the Planck scale:
\begin{equation}
\left(\frac{\tau_p}{\tau}\right)^2 \left(\frac{\bar E\times \tau}{\hbar}\right)=\left(\frac{L_p}{L}\right)^2 \left(\frac{\bar E\times \tau}{\hbar}\right)=\frac{L^2}{g^2}R_{(4)}(g,L).
\label{fluctuate}
\end{equation}
Here $g$ is the covariant metric tensor, $g^{-1}$  the contravariant metric tensor and the Riemann tensor, 
written in terms of the metric variations $\Delta g$, is
\begin{equation}
R_{(4)}(g,L) \sim \frac{g^2}{L^2} \left( \Delta\left(\Delta g (g)^{-1}\right) + \left(\Delta g (g)^{-1}\right)^2\right),
\end{equation}
where 
\begin{equation}
\left(\Delta g~ (g)^{-1}\right)^2_{\{kl,im-km,il\}} = \Gamma^n_{kl}\Gamma^p_{im} - \Gamma^n_{km}\Gamma^p_{il}
\end{equation}
are the Christoffel symbols and $\Delta\left(g^{-1}\Delta g\right)$  the second derivatives of the metric tensor with respect to the coordinates,
\begin{equation}
\partial^2_{kl}~g = \frac{g}{L^2}\Delta_k\Delta_l~g = 
\frac{g^2}{L^2} \Delta_k \left(\Delta_l g (g)^{-1}\right).
\end{equation}

The indetermination relationship 
of Eq. \eqref{fluctuate} corresponds to the presence of fluctuations of the averaged quantity of the proper energy $\bar E$ averaged over the volume $L^3$. 
If $\bar E = \Delta E^*$ and $\tau = \Delta t$, Eq. \eqref{fluctuate} becomes
\begin{equation}
\Delta E^*\times \Delta t= \left(\frac{\tau}{\tau_p}\right)^2\frac{L^2}{g^2}R_{(4)}(g,L) =\frac{1}{g^2} \left(\frac{L^2}{L_p}\right)^2 R_{(4)}(g,L) \geq\frac{1}{2}\left[1+\eta\left(\Delta E^*\right)^{2}+\ldots \right],
\label{ind1}
\end{equation}
and at Planck scales becomes
\begin{equation}
\Delta E^*\times \Delta t=\hbar \frac{L_p^2}{g^2}R_{(4)}(g,L)=\Delta\left(\Delta g (g)^{-1}\right) + \left(\Delta g (g)^{-1}\right)^2\geq\frac{1}{2}\left[1+\eta\left(\Delta E^*\right)^{2}+\ldots \right],
\label{ind2}
\end{equation}
where $\Delta E^* = \Delta E + \Delta g ~ \Lambda + g \Delta \Lambda$ (with $\Lambda$ the cosmological constant) is averaged on the volume $L^3$ of the 3D space-like hypersurface $\sigma$.
This would suggest that, starting from Einstein's equations, already at the first order, one has a background curvature and fluctuations with wavelength $\lambda=L$ that are connections between events in the spacetime due to an exchange of virtual gravitons with wavelength $\lambda$, in the classical interpretation.

A more exotic interpretation can be given with the ER = EPR scenario, 
where the two events forming the spacetime texture (or the relationship between the two \textquotedblleft particles'' of our \textquotedblleft gravitational hydrogen atom'') 
represent the connection of events in spacetime through an Einstein-Rosen (ER) wormhole supposed to be equivalent to a connection between events through Einstein-Podolski-Rosen quantum entangled states \cite{erepr}.

One notes also another important difference between the classical 
hydrogen atom of quantum mechanics \cite{key-7}
and the quantum SBH \cite{key-5}. 
In the standard hydrogen atom the nucleus and the electron are real and different
particles. In the quantum SBH here analysed the two actors described by our oscillator
behave instead as equal particles, as one can easily find described in the system of equations 
(\ref{eq: energia potenziale BH effettiva}) - (\ref{eq: energia totale effettiva}).
Thus, what we call the ``nucleus'' and the ``electron'', in this mathematical description
can be mutually exchanged without varying the actual physical properties of the system,
a quantum oscillator. 
Therefore, the quantum state described by the quantum oscillator 
results even more uncertain, more non-deterministic,
more smeared and more probabilistic than the corresponding quantum
states of the particles of the hydrogen atom. 
The results in \cite{key-5} are also in agreement with
the general conviction that quantum gravity effects become fundamental
in the presence of strong gravitational fields. In a certain sense,
the results in \cite{key-5} that have been reanalysed in this Section allow us to ``see into'' the SBH. 

Recently, the analysis in \cite{key-5} has been extended to the Reissner-Nordstrom
BH \cite{key-20}. Moreover, the stability of the solutions has been discussed, showing the existence of oscillatory regimes or exponential damping for the evolution of a small perturbation from a stable state. 

The close relationship between GUP and the emergence of a minimal length suggests a scenario that already 
seems useful for a future quantum gravity. The analysis carried out so far can be summed up by saying that there 
is a non-local core in a BH, a \textquotedblleft critical point" of the metric description in which classical concepts collapse and 
a quantum regime comes into play. In dynamic terms one can see it as the formation of a zone of high coherence \cite{Lindesay}.
A non-local state can be associated with a physical regime described by a non-switching geometry that results 
in typical oscillating processes \cite{Muthukumar,Haldar,Lin,Dadic}.

\section{Path integral formulation}
Alternatively, one can obtain the same results through a path integral approach, as expected. Let us consider the standard Einstein - Hilbert Lagrangian \cite{key-22}:
\begin{equation}
\mathcal{L}_{EH}=\frac{\sqrt{-g}R}{16\pi}.\label{eq: EH}
\end{equation}
By using the FLRW line-element (\ref{eq: FLRW}), one gets 
\begin{equation}
\mathcal{L}_{FLRW}=\dot{a}^{2}+\frac{8}{3}\pi a^{2}\rho.\label{eq: Lag FLRW}
\end{equation}
Let us rescale Eq. (\ref{eq: Lag FLRW}) as 
\begin{equation}
\mathcal{L}=\frac{1}{2}M\dot{a}^{2}+\frac{4}{3}M\pi a^{2}\rho,\label{eq: Lag rescaled}
\end{equation}
which can be written as
\begin{equation}
\mathcal{L}=\frac{M\dot{a}^{2}}{2}-V(a),\label{eq: Lagrangiana F}
\end{equation}
where 
\begin{equation}
V(a)\equiv-\frac{4}{3}M\pi a^{2}\rho.\label{eq: potenziale F}
\end{equation}
The energy function associated to the Lagrangian is 
\begin{equation}
E=\frac{\partial \mathcal{L}}{\partial\dot{a}}\dot{a}-\mathcal{L}.\label{eq: en lagr}
\end{equation}
Then, by inserting Eq. (\ref{eq: Lag rescaled}) in Eq. (\ref{eq: en lagr}),
and by using the first of Eqs. (\ref{eq: evoluzione}), one gets 
\begin{equation}
E=-\frac{M}{2},\label{eq: energia totale F}
\end{equation}
which is exactly Eq. (\ref{eq: energia totale 2}). 

Now, by taking an infinitesimal time interval $[\tau,\tau+\delta\tau]$,
the wave function $\Psi$ for the system at the time $\tau+\delta\tau$
in terms of its value at time $\tau$ is determined by the integral
equation \cite{key-23} 
\begin{equation}
\Psi\left(Q,\tau+\delta\tau\right)=\int\exp\left[i\delta\tau \mathcal{L}\left(\frac{Q-q}{\delta\tau},Q\right)\right]\Psi\left(q,\tau\right)\frac{\sqrt{g\left(q\right)}dq}{A\left(\delta\tau\right)},\label{eq: wave function F}
\end{equation}
where the transformation function $\left(q'_{\tau+\delta\tau}\equiv Q|q'_{\tau}\equiv q\right)$,
which connects the representations referring to the two different
times $\tau$ and $\tau+\delta\tau$, corresponds in the classical
theory to $\exp\left(i\delta\tau L\right)$, $\sqrt{g\left(q\right)}dq$
is the volume element in q-space and $A\left(\delta\tau\right)$ is
a suitable normalization constant, see \cite{key-23} for further details.
In accordance with Eq. (\ref{eq: wave function F}), the wave
function for the system must satisfy, for infinitesimal $\varepsilon$, the equation (where one writes $\varepsilon$
for $\delta\tau$)
\begin{equation}
\Psi\left(a,\tau+\varepsilon\right)=\int\exp\left\{i\varepsilon\left[\frac{M}{2}\left(\frac{a-y}{\varepsilon}\right)^{2}-V(a)\right]\right\} \Psi\left(y,\tau\right)\frac{dy}{A}.
\end{equation}
Now, one replaces $y=\eta+a$ in the integral, obtaining
\begin{equation}
\Psi\left(a,\tau+\varepsilon\right)=\int\exp\left\{ i\left[\frac{M}{2}\frac{\eta}{\varepsilon}^{2}-\varepsilon V(a)\right]\right\} \Psi\left(\eta+a,\tau\right)\frac{d\eta}{A}.\label{eq: evoluzione F 2}
\end{equation}
Only values of $\eta$ close to zero will contribute to the integral. Then, one
expands $\Psi\left(\eta+a,\tau\right)$ in a Taylor series around
$\eta=0$. After rearranging, one gets
\begin{equation}
\Psi\left(a,\tau+\varepsilon\right)=\frac{\exp\left(-i\varepsilon V(a)\right)}{A}\int\left[\exp\left(i\frac{M}{2}\frac{\eta}{\varepsilon}^{2}\right)\right]\left[\Psi+\eta\Psi'+\frac{\eta^{2}}{2}\Psi''+\ldots\right]d\eta,
\end{equation}
where, for the sake of simplicity, we defined $\Psi \equiv \Psi(a, \tau)$, $\Psi'(a, \tau)\equiv\frac{\partial\Psi\left(a,\tau\right)}{\partial a}$ and $\Psi''(a,\tau)\equiv \frac{\partial^{2}\Psi\left(a,\tau\right)}{\partial a^{2}}$.
From Pierces integral tables in Feynman's PhD thesis \cite{key-23} one gets 
\begin{equation}
\int_{-\infty}^{\infty}\exp\left(i\frac{M}{2}\frac{\eta}{\varepsilon}^{2}\right)d\eta=\sqrt{\frac{2\pi i\varepsilon}{M}},\label{eq: Pierces}
\end{equation}
and, by differentiating both sides with respect to $M$, one finds
\begin{equation}
\int_{-\infty}^{\infty}\exp\left(i\frac{M}{2}\frac{\eta}{\varepsilon}^{2}\right)\eta^{2}d\eta=\sqrt{\frac{2\pi i\varepsilon}{M}}\frac{\varepsilon i}{M}.\label{eq: Pierces 2}
\end{equation}
The integral with $\eta$ in the integrand is the integral of an odd function, and so it is zero. Hence one obtains
\begin{equation}
\Psi\left(a,\tau+\varepsilon\right)=\frac{1}{A}\sqrt{\frac{2\pi i\varepsilon}{M}}\int_{-\infty}^{\infty}\exp\left[-i\varepsilon V(a)\right]\left[\Psi+\frac{\varepsilon i}{2M}\Psi''+\ldots\right].\label{eq: evoluzione F 4}
\end{equation}
The left hand side of Eq. (\ref{eq: evoluzione F 4}) approaches $\Psi\left(a,\tau\right)$
for small $\varepsilon$. The equality holds if
\begin{equation}
A\left(\varepsilon\right)=\sqrt{\frac{2\pi i\varepsilon}{M}}.\label{eq: A}
\end{equation}
If one expands both sides of Eq. (\ref{eq: evoluzione F 4}) in powers
of $\varepsilon$ up to the first order, one finds
\begin{equation}
i\frac{\partial\Psi}{\partial\tau}=-\frac{1}{2M}\frac{\partial^{2}\Psi}{\partial a^{2}}+V(a)\Psi,\label{eq: Schrodinger equation F}
\end{equation}
which is exactly the Schr\"odinger equation (\ref{eq: Schrodinger equation}).
Hence, now one can remake the analysis below Eq. (\ref{eq: Schrodinger equation})
in Section I obtaining the same results.

\section{Black hole entropy}
Now, one can improve the analysis made in Section II by discussing the SBH
entropy. So, let us consider a SBH excited at the quantum level $n$;
then, by putting $A_{n}\equiv16\pi M_{n}^{2}$, the number of quanta
of area is 
\begin{equation}
N_{n}\equiv\frac{A_{n}}{\Delta A_{n}}=\frac{16\pi M_{n}^{2}}{64\pi}=\frac{1}{4}M_{n}^{2}=n,\label{eq: N n}
\end{equation}
where Eq. (\ref{eq: spettro massa BH finale}) has been used. Thus,
one finds the intriguing (but also intuitive) result that the number
of quanta of area is equal to the SBH principal quantum number. Then,
the formula of the famous Bekenstein-Hawking entropy \cite{key-47,key-28}
reads 
\begin{equation}
\left(S_{BH}\right)_{n}\equiv\frac{A_{n}}{4}=\frac{N_{n}\Delta A_{n}}{4}=16\pi n,\label{eq: Bekenstein-Hawking  n}
\end{equation}
that means that Bekenstein-Hawking is linear function of the SBH principal
quantum number, i.e. of the BH excited state. By using the quantum
tunnelling approach one can arrive to the sub-leading corrections
at third order approximation \cite{key-45} 
\begin{equation}
S_{total}=S_{BH}-\ln S_{BH}+\frac{3}{2A}+\frac{2}{A^{2}}.\label{eq: entropia totale}
\end{equation}
Thus, by using Eq. (\ref{eq: Bekenstein-Hawking  n}) one gets 
\begin{equation}
\left(S_{total}\right)_{n}=16\pi n-\ln\left(16\pi n\right)+\frac{3}{128\pi n}+\frac{2}{\left(64\pi n\right)^{2}}.\label{eq: entropia totale n}
\end{equation}
One stresses that in the above analysis it has been implicitly assumed
that the coefficient of Bekenstein-Hawking entropy has its traditional
value $\frac{1}{4}.$ On the other hand, following the analysis in
\cite{key-46}, one can show that the mass and energy spectra given
by Eqs. (\ref{eq: spettro massa BH finale}) and (\ref{eq: BH energy levels finale.})
imply that the value of such a coefficient is different. For the sake of completeness, here
one must recall that the use of the microcanonical ensemble for describing Hawking radiation and BH evaporation was advocated by Hawking himself in \cite{key-76} and later on revived in \cite{key-77}, way before \cite{key-46}. 

One starts to recall that the microcanonical ensemble is the proper
ensemble to describe BHs which are not in thermodynamic equilibrium,
such as radiating BHs \cite{key-46}. This choice of ensemble eliminates
the problems, i.e., negative specific heat and loss of unitarity,
encountered when the canonical ensemble is used \cite{key-46}.

Now, let energy eigenvalues $E_{n}$ with multiplicities $\nu(n)$ be \cite{key-46}
\begin{equation}
E_{n}=\gamma\sqrt{n},\hspace{0.5cm}\nu(n)=g^{n},\hspace{0.3cm}n=1,2,\ldots\label{eq: autovalori con moltiplicit=0000E0}
\end{equation}
where $\gamma=-\frac{1}{2}$ from Eq. (\ref{eq: BH energy levels finale.})
and $g>1$. The microcanonical partition function $\Omega(E)$ is
equal to the number of eigenvalues below the energy $E$. Hence 
\begin{equation}
\sum_{n=1}^{n(E)}g^{n}=\frac{g^{n(E)}-1}{g-1},\label{eq: somme}
\end{equation}
where \cite{key-46}
\begin{equation}
n(E)=\frac{E^{2}}{\gamma^{2}}=4E^{2}.\label{eq: n di E}
\end{equation}
One considers Bekenstein's original intuition that the BH entropy
can be written as \cite{key-47}
\begin{equation}
S_{BH}=f(A)=cA,\label{entropy}
\end{equation}
where $A$ is the area of the event horizon. The aim is to determine
the constant $c$ starting from the analysis developed in previous
Sections. Starting from Eq. ($\ref{entropy}$), it is natural to assume
an $\exp(cA)$-fold degeneracy in the possible BH mass eigenvalues
$M_{n}$ \cite{key-46}. This assumption of degeneracy is justified
because entropy, in general, can be understood as a logarithm of the
number of microstates corresponding to the same macrostate \cite{key-46}.
Since for a SBH with mass $M$, $A=16\pi M^{2}$, one is prompted
to define $\nu(M_{n})$ as the number of degenerate states corresponding
to the same mass eigenvalue $M_{n}$ such that \cite{key-46} 
\begin{equation}
\nu(M_{n})=\nu(n)=\exp{(16c\pi M_{n}^{2})}.\label{eq: degenerazione}
\end{equation}
Considering Eq. (\ref{eq: spettro massa BH finale}) one can write
\cite{key-46} 
\begin{equation}
g=\exp{(64c\pi)}.
\end{equation}
Now, one comes back to the microcanonical system. As in \cite{key-46}, to find the thermodynamics, one has to compute
\begin{equation}
\omega=\frac{\partial\Omega}{\partial E}=\frac{\log(g)}{g-1}8Eg^{4E^{2}}.
\end{equation}
Then, the microcanonical temperature is given by
\begin{equation}
k_{B}T=\frac{\Omega}{\omega}=\frac{g^{4E^{2}}-1}{g^{4E^{2}}}\frac{1}{8E\log(g)}.
\end{equation}
Its inverse is equal to 
\begin{equation}
\beta=8E\log(g)\left[1+\mathcal{O}\left(g^{-4E^{2}}\right)\right].\label{eq: ordine}
\end{equation}
From Section II one has $M=|4E|$. Then 
\begin{equation}
\beta=2M\log(g)=128Mc\pi,
\end{equation}
up to the exponentially small corrections in Eq. (\ref{eq: ordine}).
By imposing $\beta=\beta_{H}=8\pi M$ (in Planck units),
one finds 
\begin{equation}
c=\frac{1}{16}.
\end{equation}
Thus, from Eq. ($\ref{entropy}$), one gets 
\begin{equation}
S_{BH}=\frac{A}{16}.
\end{equation} 

Hence, it has been shown that the approach to BH quantum physics that
has been developed in previous Sections implies that the coefficient
of the Bekenstein-Hawking entropy is a quarter of its traditional
value. Then, Eqs. (\ref{eq: Bekenstein-Hawking  n}) and (\ref{eq: entropia totale n}) becomes
\begin{equation}
\left(S_{BH}\right)_{n}\equiv\frac{A_{n}}{16}=\frac{N_{n}\Delta A_{n}}{16}=4\pi n,\label{eq: Bekenstein-Hawking n  rivista}
\end{equation}
and 
\begin{equation}
\left(S_{total}\right)_{n}=4\pi n-\ln\left(4\pi n\right)+\frac{3}{32\pi n}+\frac{2}{\left(16\pi n\right)^{2}},\label{eq: entropia totale n rivista}
\end{equation}
respectively.

\section{Consistence with the semi-classical Bohr-like approach to black hole
quantum physics}
The Bohr-like approach to BH quantum physics has been previously developed
by one of the Authors (CC) (see \cite{key-8,key-9} for further details) and reviewed in
\cite{key-21}, starting from the pioneering works \cite{key-24,key-25}.
This approach is founded on the concept of \emph{BH effective state, }which
has been partially introduced in Section II of this paper through the
definitions of the BH\emph{ }effective mass and effective horizon,
and the natural correspondence between Hawking radiation and BH quasi-normal
modes (QNMs). Such a correspondence
allows indeed to naturally interpret BH QNMs as quantum levels in
a semi-classical approach. This
is an approach to BH quantum physics somewhat similar to the historical
semi-classical approach to the structure of a hydrogen atom introduced
by Bohr in 1913 \cite{key-26,key-27}; in a certain sense, QNMs represent
the ``electron'' which jumps from a level
to another one and the absolute values of the QNMs frequencies, ``triggered''
by emissions (Hawking radiation) and absorption of external particles,
represent the energy ``shells'' of the GHA. After shortly reviewing the results in \cite{key-8,key-9,key-21,key-24,key-25},
the approach will be refined through an important rescaling of the
quantum levels and some further modify which will take into account
the real physical behaviors of the SBH. In that way, it will be shown
that this approach to BH physics
will be completely consistent with the full quantum treatment of previous
Sections. This will give a remarkable physical insight; to clarify
this point, let us again consider the analogy between the potential
energy of a hydrogen atom, given by Eq. (\ref{eq: energia potenziale atomo idrogeno}),
and the potential energy of the GHA given by Eq. (\ref{eq: energia potenziale BH effettiva}).
Eq. (\ref{eq: energia potenziale atomo idrogeno}) represents
the interaction between the nucleus of the hydrogen atom, having a
charge $e$ and the electron, having a charge $-e$ while Eq. (\ref{eq: energia potenziale BH effettiva})
represents the interaction between the nucleus of the GHA, i.e., the BH, having an effective, dynamical mass
$M_{E}$, and another, mysterious, particle, i.e., the ``electron''
of the GHA having again an effective,
dynamical mass $M_{E}$.\textbf{ }The Bohr-like approach to BH quantum
physics shows that the ``electron states'' of the BH are exactly the BH QNMs \textquotedblleft triggered\textquotedblright{}
by emissions of Hawking quanta and absorption of external particles.
Remarkably, the QNM jumping from a level to another one has been indeed
interpreted in terms of a particle quantized on a circle, which is analogous to the electron traveling in circular orbits around
the hydrogen nucleus, similar in structure to the solar system, of
Bohr's semi-classical model of the hydrogen atom \cite{key-26,key-27}.

Let us recall that Hawking radiation
is today largely analysed through the tunneling mechanism (see \cite{key-29,key-30,key-31,key-32,key-33,key-34,key-35} and references within); by considering a classically stable object, if one sees that it becomes quantum-mechanically
unstable, suspecting tunneling is natural. Then, particles
creation by BH can be described as tunneling generated
by vacuum fluctuations near the BH horizon \cite{key-29,key-30,key-31,key-32,key-33,key-34,key-35}.
If a virtual particle pair originates just inside the BH horizon,
the virtual particle with positive energy can tunnel outside the BH
becoming a real particle. The same happens when a virtual particle
pair originates just outside the horizon. Then, the particle with
negative energy can tunnel inwards. It results that the particle with
negative energy is absorbed by the BH in both cases. Consequently,
the BH mass decreases and the particle with positive energy moves
towards infinity. This is exactly the mechanism of emission of Hawking
radiation. Working in strictly thermal approximation, the probability
of emission of Hawking quanta is \cite{key-28,key-29} 
\begin{equation}
\Gamma\sim\exp\left(-\frac{\omega}{T_{H}}\right),\label{eq: hawking probability}
\end{equation}
where $\omega$ is the energy-frequency of the emitted particle
and $T_{H}\equiv\frac{1}{8\pi M}$ is the Hawking temperature. But,
if one considers the energy conservation, that means the
BH contraction which enables a varying BH geometry, one obtains the following correction \cite{key-29}:
\begin{equation}
\Gamma\sim\exp\left[-\frac{\omega}{T_{H}}\left(1-\frac{\omega}{2M}\right)\right],
\end{equation}
and so
\begin{equation}\label{eq: Parikh Correction}
\Gamma=\alpha\exp\left[-\frac{\omega}{T_{H}}\left(1-\frac{\omega}{2M}\right)\right], 
\end{equation}
where $\alpha\sim1$ and one gets the additional term $\frac{\omega}{2M}\:$.
The tunnelling picture can be improved by showing that the probability
of emission (94) is really associated to
the two distributions \cite{key-8,key-34} 
\begin{align}
\braket{N}_{boson}&=\frac{1}{\exp\left[4\pi\left(2M-\omega\right)\omega\right]-1}\label{eq: 95}\\
\braket{N}_{fermion}&=\frac{1}{\exp\left[4\pi\left(2M-\omega\right)\omega\right]+1}\label{eq: 96},
\end{align}
for bosons and fermions, respectively, which are \emph{non} strictly thermal. 

Now, one considers Dirac delta perturbations representing subsequent
absorptions of particles with negative energies \cite{key-8,key-9}.
Those perturbations are associated to the emission of Hawking radiation
\cite{key-8,key-9}; BH response to perturbations are QNMs \cite{key-8,key-9,key-21,key-24,key-25}.
These are frequencies of radial spin-$j$ perturbations which obeys
a time independent Schr\"odinger-like equation, see references \cite{key-8,key-9,key-21,key-24,key-25} for further details.
QNMs represent indeed BH modes of energy dissipation having complex
frequency \cite{key-8,key-9,key-21,key-24,key-25}. The first attempt
to model the semi-classical BH in terms of BH QNMs can be found in reference
\cite{key-36}. For large values of the principal quantum number $n$,
with $n=1,2,\ldots$, QNMs result independent of both the spin and the
angular momentum quantum numbers \cite{key-8,key-9,key-21,key-24,key-25}.
This is in agreement with Bohr's Correspondence Principle \cite{key-37},
which is one of the most important principle concerning the approach
of quantum theory through semi-classical approximation. It states
that \textquotedblleft trans\emph{ition frequencies at large quantum
numbers should equal classical oscillation frequencies}\textquotedblright{}
\cite{key-37}. In other words, Bohr's Correspondence Principle permits
a precise semi-classical analysis of quantum phenomena when the values
of the principal quantum number $n$ are large. In our case, it enables
to realize a precise analysis of excited BHs \cite{key-8,key-9,key-21,key-24,key-25}.
By considering this important principle, it has been indeed shown
that QNMs can supply information on area quantization because QNMs
perturbations can be associated to absorption of external particles
\cite{key-38}. The analysis in \cite{key-38} was then improved in \cite{key-39}
by solving some important issues. In any case, an important problem was that,
being QNMs \emph{countable} frequencies, the \emph{continuous} character
of Hawking radiation did not originally allow to interpret QNMs in
terms of emitted quanta \cite{key-40}. This avoided to associate
QNMs to Hawking radiation \cite{key-40}. More recently, it has been
observed (see \cite{key-8,key-9,key-21,key-24,key-25})
that an important consequence of the non-thermal spectrum in \cite{key-29}
is the \emph{countable} character of subsequent emissions of Hawking
particles. This generates a natural correspondence between Hawking
radiation and BH QNMs which allows to interpret QNMs also in terms
of emitted quanta. In fact,
the key point is that Dirac delta perturbations arising from discrete
subsequent absorptions of particles with negative energies, which
are associated to Hawking emissions, generates indeed a BH back reaction
in terms of QNMs. Thus, the
BH contraction arising from energy conservation has not to be considered
as being a ``one shot process''. It consists instead in oscillations
of the BH horizon which are the BH QNMs. It is indeed well known that
the correspondence between emitted radiation and proper oscillation
of the emitting body is considered a fundamental behavior of every
radiation process in Nature. In other words, one can extend the analysis
in \cite{key-38,key-39} by considering QNMs in terms of quantum levels
also for emitted energies \cite{key-8,key-9,key-21,key-24,key-25}.
This is the fundamental idea on which the Bohr-like approach to BH
quantum physics is founded \cite{key-8,key-9,key-21,key-24,key-25}.

Let us shortly review how the Bohr-like approach works. One introduces
the \emph{effective temperature}
\begin{equation}
T_{E}(\omega)\equiv\frac{2M}{2M-\omega}T_{H}=\frac{1}{4\pi(2M-\omega)},\label{eq: Corda Temperature}
\end{equation}
and rewrites Eq. (\ref{eq: Corda Temperature}) in a Boltzmann-like
form similar to Eq. (\ref{eq: hawking probability}):
\begin{equation}
\Gamma=\alpha\exp[-\beta_{E}(\omega)\omega]=\alpha\exp\left(-\frac{\omega}{T_{E}(\omega)}\right),\label{eq: Corda Probability}
\end{equation}
where $\exp[-\beta_{E}(\omega)\omega]$ is the \emph{effective Boltzmann
factor,} with $\beta_{E}(\omega)\equiv\frac{1}{T_{E}(\omega)}$.
Hence, $T_{E}(\omega)$ replaces $T_{H}$ in the equation of the probability
of emission. It is important
to recall that in various fields of Science the deviation from the
thermal spectrum of an emitting body can be taken into account through
the introduction of an effective temperature representing the temperature
of a black body that emits the same total amount of radiation\emph{.}
The concept of effective temperature in BH physics has been introduced
in the pioneering works \cite{key-24,key-25} and then used in \cite{key-8,key-9,key-20,key-21}
and in other works (references within \cite{key-21}). The \emph{effective temperature} $T_{E}(\omega)$
depends on the energy-frequency of the emitted radiation and the deviation
of the BH radiation spectrum from the strictly thermal feature can
be quantified through the ratio
\begin{equation}
\frac{T_{E}(\omega)}{T_{H}}=\frac{2M}{2M-\omega}.
\end{equation} 
It was exactly the introduction
of the effective temperature which permitted the introduction of the
other effective quantities (see Eq. (\ref{eq: effective quantities absorption}).
Thus, in order to finalize the discussion on the \emph{BH effective
state} of Section II, one recalls that the effective temperature \emph{$T_{E}\:$
}is the inverse of the average value of the inverses of the initial
and final Hawking temperatures. (\emph{Before} the emission we have
\begin{equation}
T^{\text{i}}_{H}=\frac{1}{8\pi M},
\end{equation}
and \emph{after} the emission
\begin{equation}
T^{\text{f}}_{H}=\frac{1}{8\pi(M-\omega)}.
\end{equation}

As stated before, by considering an excited SBH, which means a SBH having large values
of the principal quantum number $n,$ then, the QNMs expression results
independent on the angular momentum quantum number,
in full agreement with the above cited Bohr's Correspondence Principle.
Then, if one wants to take into account the non-strictly thermal behavior
of the radiation spectrum, one replaces \emph{$T_{H}$} with \emph{$T_{E}\:$}
in the standard, strictly thermal, equation for the QNMs:
\begin{equation}\label{eq: quasinormal modes corrected}
\omega_{n}=a+ib+2\pi in T_{E}(|\omega_{n}|) \simeq2\pi in T_{E}(|\omega_{n}|)=\frac{in}{4M-2|\omega_{n}|}=\frac{in}{4M_{E}},
\end{equation}
where $a=\ln(3) T_{E}(|\omega_{n}|), b=\pi T_{E}(|\omega_{n}|)$
for $j=0,2$ (scalar and gravitational perturbations), $a=0,\;b=0$
for $j=1$ (vector perturbations) and $a=0,\;b=\pi T_{E}(|\omega_{n}|)$
for half-integer values of $j$. In Eq. (\ref{eq: quasinormal modes corrected}), $M_{E}$ is the BH effective mass
(see also Eq. (\ref{eq: effective quantities absorption})), where
the minus sign has been taken into account since an emission is considered.
On the other hand, as $a,b\ll|2\pi inT_{E}(|\omega_{n}|)|$ for large
$n$, the leading term in the imaginary part of the complex frequencies
well approximates the QNMs (\ref{eq: quasinormal modes corrected}).
Therefore, the spin content of the perturbation does not influence
the leading asymptotic behavior of $|\omega_{n}|$. This is again consistent with the Bohr's Correspondence Principle.
At order of leading asymptotic behavior the solution of Eq. (\ref{eq: quasinormal modes corrected})
is \cite{key-8,key-9,key-21,key-24,key-25}
\begin{equation}
|\omega_{n}|=M-\sqrt{M^{2}-\frac{n}{2}}.\label{eq: radice fisica}
\end{equation}
In \cite{key-8,key-9,key-21,key-24,key-25}, $|\omega_{n}|\:$ has
been interpreted like the total energy emitted for a BH excited at
a level $n$. Now, considering an emission from a BH at rest
to a state with large $n=n_{1}$, and using Eq. (\ref{eq: radice fisica}),
one sees that the BH mass changes from $M$ to
\begin{equation}
M_{n_{1}}\equiv M-|\omega_{n_{1}}|=\sqrt{M^{2}-\frac{n_{1}}{2}}.\label{eq: me-1}
\end{equation}
If we consider a subsequent emission from the quantum level $n=n_{1}$
to the quantum level $n=n_{2}$, with $n_{2}>n_{1}$, then the BH
mass changes again from $M_{n_{1}}$ to
\begin{equation}\label{eq: me}
M_{n_{2}}\equiv M_{n_{1}}-\Delta E_{n_{1}\rightarrow n_{2}}=M-|\omega_{n_{2}}|=\sqrt{M^{2}-\frac{n_{2}}{2}},
\end{equation}
with
\begin{equation}\label{eq: jump}
\Delta E_{n_{1}\rightarrow n_{2}}\equiv|\omega_{n_{2}}|-|\omega_{n_{1}}| =M_{n_{1}}-M_{n_{2}}=\sqrt{M^{2}-\frac{n_{1}}{2}}-\sqrt{M^{2}-\frac{n_{2}}{2}},
\end{equation}
which is the quantum jump between the two BH energy levels due to the emission
of a particle having frequency $\Delta E_{n_{1}\rightarrow n_{2}}$.
Hence, the quantum jump implies the emission of a discrete amount
of energy.
In the case of Bohr's hydrogen atom \cite{key-26,key-27}, the electron only gains
and loses energy when it jumps from one allowed energy shell to another.
It can absorb or emit radiation having an energy difference of the
levels governed by the Planck relation (in standard units) $E=hf$,
being $\:h\:$ the Planck constant and $f\:$ the transition frequency.
In the Bohr-like approach to BH physics, QNMs can only gain and lose
energy by jumping from one allowed energy shell to another.
Thus, they absorb or emit radiation (emitted radiation is given by
Hawking quanta) with an energy difference of the levels given by to
Eq. (\ref{eq: jump}). Remarkably, one can interpret Eq. (\ref{eq: radice fisica})
in terms of a particle, the ``electron'', quantized on a circle
of length \cite{key-8,key-9,key-21,key-24,key-25} 
\begin{equation}
L=\frac{1}{T_{E}(|\omega_{n}|)}=4\pi\left(M+\sqrt{M^{2}-\frac{n}{2}}\right).\label{eq: lunghezza cerchio}
\end{equation}
This is analogous to the electron traveling in circular orbits around
the hydrogen nucleus.

Now, the semi-classical Bohr-like approach to BH physics will be refined
through some simple, but important physical observation, and this
will allow us to find a perfect consistence the approaches which have been analysed in Sections II and III. Let us observe that the Bohr-like approach
has been developed starting from the assumption of a SBH at rest which
starts to emit Hawking radiation. But the key point is that a BH can never be at rest. Let us clarify this point. After the BH formation,
there is a ``settlement'' phase due to ``gravitational
recoil''. As one is considering SBHs, which arise from a
radial gravitational collapse having spherical symmetry, there is
no gravitational wave (GW) emission in such a ``settlement''
phase. In fact, in general relativity, GWs are quadrupole waves and
a perfect spherical symmetric source cannot emit GWs \cite{key-4}.
On the other hand, from the quantum point of view, one can consider
the BH formation as subsequent absorptions of separated particles
starting from a BH having null mass. In that
case, the approach in \cite{key-38,key-39} implies that particles
absorptions by the BH generates perturbations and that the BH back
reaction to such perturbations is given by QNMs. In other words, the
``settlement'' phase after the BH formation
is completely described by the BH QNMs ``triggered'' by the absorption
of particles. When will this
phase end? In order to answer this question one has to recall that
a BH is not immerged in vacuum. Instead, it is immerged in the thermal
bath of the Cosmic Background Radiation (CBR). Thus, it will continue
to absorb CBR photons (and other potential external particles) until
the CBR temperature will be higher than the BH Hawking temperature.
In other words, the ``settlement'' phase
will end when the BH Hawking temperature will become higher than the
CBR temperature. But, at that point, the BH will start to emit Hawking
radiation and the BH QNMs will be ``triggered'' by the absorption
of Hawking quanta having negative energies. The process will continue
again and again, until the Planck distance and the Planck mass are
approached. At that point, as it has been stressed in Section II, the
GUP prevents the BH from its total evaporation by stopping Hawking's
evaporation process in exactly the same way that the HUP prevents
the hydrogen atom from total collapse \cite{key-17}. Therefore, the BH
arrives at the minimum energy level which is compatible with the GUP;
such an energy level corresponds to our Eq. (\ref{eq: energia minima}).

From the above discussion one argues that the only BH at rest is the BH having null mass and one must consider absorptions
instead of emissions. This means that one must replace $M\rightarrow0$ and $-\omega_{n}\rightarrow M$
in the right hand side of Eq. (\ref{eq: quasinormal modes corrected}).
One also recalls that
the principal quantum number increases for absorptions instead of
emissions. Thus, one obtains 
\begin{equation}
\omega_{n}=\frac{in}{2M}=\frac{in}{4M_{E}}\label{eq: quasinormal modes finale},
\end{equation}
and, again, one gets
\begin{equation}
M_{E}=\frac{M}{2}.  
\end{equation}
One gets complete consistence with the two-particle quantum system
of Eqs. (\ref{eq: energia potenziale BH effettiva}) and (\ref{eq: Schrodinger equation BH effettiva}) by interpreting the quasi-normal frequency (\ref{eq: quasinormal modes finale}) as being
a particle, the ``electron'', quantized on a circle. 
But it has been previously observed that, in such a two-particle quantum
system, the two particles are equal and can be mutually exchanged
without varying the physical properties of the system. This means
that one can assign to the QNMs in Eq. (\ref{eq: quasinormal modes finale})
the half of the total energy of the two-particle quantum
system, obtaining 
\begin{equation}
\omega_{n}=\frac{E_{n}}{2}=\frac{in}{2M}=\frac{in}{4M_{E}}\label{eq: mezza energia}
\end{equation}
Let us also remark that the total BH energy $E=-\frac{M}{2}$
\emph{is }negative and different from the BH inert mass $M.$ Thus,
by using Eq. (\ref{eq: energia totale effettiva}), one gets from Eq.
(\ref{eq: mezza energia}) 
\begin{equation}
\frac{M_{E}}{4}=\left|\frac{in}{4M_{E}}\right|,\label{eq: un quarto di massa effettiva}
\end{equation}
that is
\begin{equation}
M_{n}=2\sqrt{n}.\label{eq: spettro massa BH finale QNMs}
\end{equation}
Remarkably, Eq. (\ref{eq: spettro massa BH finale QNMs}) is exactly the same
Eq. (\ref{eq: spettro massa BH finale}). Thus, one can use again Eq. (\ref{eq: energia totale effettiva}) in order to re-obtain
Eq. (\ref{eq: BH energy levels finale.}). 

In complete consistence with Section 2, Eq. (\ref{eq: mezza energia}) is interpreted in terms of a particle, the ``electron'', quantized on a circle of length
\begin{equation}
L=\frac{1}{T_{E}}=8\pi M_{E}=8\pi\sqrt{n},\label{eq: lunghezza cerchio corretta}
\end{equation}
which is equal to Eq.  (\ref{eq: lunghezza cerchio corretta 1}) which was derived in Section 2 through a full quantum analysis. The semi-classical physical interpretation
of the QNMs in Eq. (\ref{eq: quasinormal modes finale}) is of a collection of damped harmonic degrees of freedom, which releases an intuitive physical picture of a BH as a whole \cite{key-39}. The
larger is $\left|\omega_{n}\right|,$ the shorter is the lifetime,
as one expectes from physical intuition \cite{key-39}. Thus, the
increasing of absorptions ``triggers'' shorter and shorter lived
modes, while subsequent emissions of Hawking quanta ``trigger''
longer and longer lived modes. 

It has been shown that different approaches give the same
results concerning the BH mass and energy spectra. It also means that one gets different
pictures concerning the SBH. In the classical framework of Einstein's
general relativity, a BH is a ``dead object'' with inert mass. It
is a definitive prison where anything that enters cannot escape. This
means that it can only become more massive and bigger with time. In
the semi-classical approximation, Hawking has shown, in one of his most famous
papers \cite{key-28}, that BHs may actually emit quanta. The emission
of Hawking radiation implies that BHs lose energy and decrease their
mass until eventually evaporating. Hawking's semi-classical picture
can be refined through the Bohr-like approach that has been developed
in this Section. The ``electron'', which is given by
the BH QNMs ``triggered'' by the absorptions of external particles
and/or by the emissions of Hawking quanta, travels in circular orbits
around the ``nucleus''. The circular orbits become more
external in the case of absorptions and more internal in the case
of emissions, see Eq. (\ref{eq: lunghezza cerchio corretta}).
In other words, the ``electron'' jumps from an orbit to another
one due to absorptions and/or emissions.
The full quantum picture discussed in Section 2 is completely consistent with the
Bohr-like approach analysed in this Section. In particular it permitted to  obtain an interesting quantum representation of the Schwarzschild BH ground state at the Planck scale which also concerns the interpretation of the BH origin in terms of evolving particles pair.

\section{Hawking radiation and quantum structure of a black hole}

It is interesting to make some considerations on the spacing of the
mass levels. In the case of an absorption, from Eq. (\ref{eq: spettro massa BH finale}), one gets
for two neighboring levels
\begin{equation}\label{eq: variazione massa}
\Delta M_{n\rightarrow n+1}=2\left(\sqrt{n+1}-\sqrt{n}\right)=\frac{2\left(n+1-n\right)}{\left(\sqrt{n+1}+\sqrt{n}\right)}=\frac{2}{\left(\sqrt{n+1}+\sqrt{n}\right)},
\end{equation}
which means that the spacing of the mass levels decreases with increasing
$n$, and that
\begin{equation}
\lim_{n\rightarrow\infty}\Delta M_{n\rightarrow n+1}=0.   
\end{equation}
For large $n$ (highly excited BHs), one gets $\sqrt{n}\simeq\sqrt{n+1.}$
Thus,
\begin{equation}
\Delta M_{n\rightarrow n+1}\simeq\frac{1}{\sqrt{n}}.\label{eq: variazione massa grande n}
\end{equation}
If one considers an astrophysical BH having mass of the order of ten solar
masses, from Eq. (\ref{eq: spettro massa BH finale}), one gets $\sqrt{n}\sim10^{38}$, which
means that $\Delta M_{n\rightarrow n+1}\simeq10^{-38}m_{P}.$ This
is an infinitesimal quantity, and this implies that, considering
astrophysical BHs, on one hand their mass spectrum becomes practically
continuous and, on the other hand, that the emission of Hawking radiation will be practically thermal. 
This is consistent with the original semi-classical computation of Hawking that can be found in \cite{key-28} and it is interesting for the following reason. In \cite{key-14} Bekenstein and Mukhanov claimed that a modification of the Hawking radiance spectrum should be necessary in order to take into account that BHs should have a discrete mass spectrum with concomitant line emission. Instead, here it has been shown that such a modification is not strictly necessary because the spacing of the levels become infinitesimal for large $n$. Thus, one can also set $M_{n}\simeq M_{n+1}\equiv M.$
Thus, from Eq. (\ref{eq: variazione massa grande n}), one finds a
minimum energy for the emissions of Hawking quanta given by 
\begin{equation}
E_{min}\simeq\frac{2}{M}=16\pi T_{H}.\label{eq: energia minima-2}
\end{equation}
One of the most interesting results we find is that the quantization of the area is compatible with the thermality of the spectrum 
since, for large values of the main quantum number, the variation of the radius of the black hole corresponding  to a Hawking emission is infinitesimal.
\\
Another important issue is that Eq. (\ref{eq: variazione massa grande n}) is the same as the one at the foundation of the interesting approach of Dvali and Gomez's to the quantum description of BHs \cite{key-79}. The approach in \cite{key-79} seems indeed consistent with the results of this paper (for instance, information is also not lost during the evaporation by construction). 
\\
Starting from these considerations, let us consider an interesting analogy with loop quantum
gravity (LQG). One observes that the results derived in previous Sections seem
consistent with the idea that the Hawking radiation spectrum should
be discrete if one quantizes the area spectrum in a way that the allowed
area is the integer multiples of a single unit area, as it has been
originally suggested in \cite{key-14}. On the other hand, the Hawking
radiation spectrum seems to be continuous in LQG if the area spectrum
is quantized in such a way that there is not only a single unit area
\cite{key-48}. But, in \cite{key-49}, by assuming the locality of photon emission in a BH, the Author suggested that the Hawking radiation spectrum is generally countable
in the LQG framework even if the allowed area is not
simply the integer multiples of a single unit area.
This result arises from the selection rule for quantum
BHs. Such an analysis shows that the Hawking radiation spectrum is
truncated below a certain frequency and hence there is a minimum energy
of an emitted particle \cite{key-49}:
\begin{equation}
E_{min}\approx\beta T_{H}=\frac{\beta}{8\pi M}.\label{eq: energia minima-1}
\end{equation}
Eq. (\ref{eq: energia minima-1}) is consistent with the result
of Eq. (\ref{eq: energia minima-2}) for $\beta=16\pi.$ 

Now, define as $n_{max}$ the maximum value of the BH principal
quantum number that corresponds to the instant when the BH temperature
equales the temperature of the surrounding CBR. After such an instant,
the BH stops to absorb CBR photons and starts to emit Hawking radiation.
Thus, from Eq. (\ref{eq: spettro massa BH finale}), one gets
\begin{equation}
M_{n_{max}}\equiv2\sqrt{n_{max}}\Longrightarrow n_{max}=\frac{M_{n_{max}}^{2}}{4},\label{eq: massa massima}
\end{equation}
and sees that, for example, in the case of an astrophysical BH having a mass
of the order of ten solar masses, it is $n_{max}\sim10^{76}$.

Now, one can compute the pre-factor $\alpha$ in Eq. (\ref{eq: Parikh Correction})
by refining the analysis in \cite{key-8} in the light of the results
of previous Sections. Let us focus on Eqs. (93) and (94). In this
kind of leading order tunneling calculations, the exponent is indeed
due to the classical action and the pre-factor is a correction of the order of the Planck constant. Then, in the case of emission of
Hawking radiation, the variation of the Bekenstein-Hawking entropy
\cite{key-8,key-29} 
\begin{equation}
\Gamma=\alpha\exp\Delta S_{BH}=\alpha\exp\left[-\frac{\omega}{T_{H}}\left(1-\frac{\omega}{2M}\right)\right],\label{eq: variation BH entropy}
\end{equation}
is the order of unity for an emitted particle having energy of the order of Hawking
temperature. Consequently, the exponent appearing in the right hand side of
Eqs. (\ref{eq: Parikh Correction}) and (\ref{eq: variation BH entropy})
is the order of unity. Thus, one can ask what is the real meaning of such
a scaling if the pre-factor is unknown. By refining the
analyses in \cite{key-8,key-9}, one shows that, fixed two quantum levels $m$ and $n$,
the mass-energy $-\Delta M_{m\rightarrow n}\equiv\omega_{m\rightarrow n}$
emitted in an arbitrary transition $m\rightarrow n$, with $n<m$,
is proportional to the effective temperature associated to the transition,
and that the constant of proportionality depends only on the difference
$m-n$. Setting \cite{key-8,key-9} 
\begin{equation}
\omega_{m\rightarrow n}=M_{m}-M_{n}=K\left[T_{E}\right]_{m\rightarrow n},\label{eq: differenza radici fisiche}
\end{equation}
and considering that, from Eq. (\ref{eq: spettro massa BH finale}),
it is
\begin{equation}
M_{m}=2\sqrt{m},\hspace{0.1cm}M_{n}=2\sqrt{n},\label{eq: masse}
\end{equation}
respectively, then one can see if there are values of the constant
$K$ for which Eq. (\ref{eq: differenza radici fisiche}) is satisfied.
As the effective temperature is the inverse of the
average value of the inverses of the initial and final Hawking temperatures, it is \cite{key-8,key-9}
\begin{equation}
\left[T_{E}\right]_{m\rightarrow n}=\frac{1}{4\pi\left(M_{m}+M_{n}\right)},\label{eq: temperatura efficace di transizione}
\end{equation}
which implies
\begin{equation}
M_{m}^{2}-M_{n}^{2}=\frac{K}{4\pi}.\label{eq: differenza radici fisiche 2}
\end{equation}
By using Eqs. (\ref{eq: masse}), Eq. (\ref{eq: differenza radici fisiche 2})
becomes 
\begin{equation}
4\left(m-n\right)=\frac{K}{4\pi},\label{eq: K solved}
\end{equation}
From last expression and from Eq. (\ref{eq: differenza radici fisiche}) one deduces that $K=16\pi\left(m-n\right).$ Hence, one finds 
\begin{equation}
\omega_{m\rightarrow n}=M_{m}-M_{n}=16\pi\left(m-n\right)\left[T_{E}\right]_{m\rightarrow n}.\label{eq: differenza radici fisiche finale}
\end{equation}
Using Eq. (\ref{eq: Corda Probability}), the probability of emission
between the two levels $m$ and $n$ can be written as
\begin{equation}
\Gamma_{m\rightarrow n}=\alpha\exp\left\{-\frac{\Delta E_{m\rightarrow n}}{\left[T_{E}\right]_{m\rightarrow n}}\right\}=\alpha\exp\left[16\pi\left(n-m\right)\right].\label{eq: Corda Probability Intriguing}
\end{equation}
Thus, the probability of emission between two arbitrary SBH quantum
levels characterized by the two principal quantum numbers $m$ and
$n$ scales like $\exp\left[16\pi\left(n-m\right)\right].$ In particular,
for $n=m-1$, the probability of emission has its maximum value $\sim\exp(-16\pi)$.
This means that the probability is maximum for two adjacent levels,
as one intuitively expects. 

Now, in order to compute the pre-factor $\alpha,$ one assumes the
unitarity of BH evaporation \cite{key-8}. Let us remark that in the next Section the BH
information paradox will be solved without using the results that
it will be obtained hereafter in this Section. This will imply the
unitarity of BH evaporation by confirming the correctness of the assumption
that one makes here. One also recalls that the majority of researchers
today think that BH evaporation is unitary. On the other hand, results
in previous Sections showed the SBH in terms of a well defined quantum
mechanical system, having an ordered, discrete quantum spectrum. This
seems consistent with the unitarity of BH evaporation. Such a unitarity
implies that, for a generic BH principal quantum
number $m$, one must have 
\begin{equation}
\sum_{n=1}^{m}\Gamma_{m\rightarrow n}=1,\label{eq: sommatoria}
\end{equation}
where the sum in the last expression has been stopped at $n=1$
because, as it has been previously stressed, the GUP prevents a BH
from its total evaporation by stopping the evaporation process at
the Planck scale, while in \cite{key-5} and in Section I of this
paper it has been shown that the Planck scale is reached at $n=1$
(BH ground state). One also notes that $n=m$ corresponds to the probability
that the BH does not emit \cite{key-8}. Now, using Eq. (\ref{eq: Corda Probability Intriguing}),
one gets 
\begin{equation}
\sum_{n=1}^{m}=\alpha\exp\left[-16\pi\left(m-n\right)\right]=1.\label{eq: sommatoria 2}
\end{equation}
Setting $k=m-n$ and $\exp\left(-16\pi\right)=X$, Eq. (\ref{eq: sommatoria 2})
becomes 
\begin{equation}
\alpha\sum_{k=0}^{m-1}X^{k}=1.\label{eq: normalizzazione 2}
\end{equation}
This sum is the k-th partial
sum of the geometric series and can be solved as
\begin{equation}
\sum_{k=0}^{m-1}X^{k}=\frac{1-X^{m}}{1-X}.\label{eq: serie geometrica}
\end{equation}
Thus, one gets 
\begin{equation}
\alpha\frac{1-X^{m}}{1-X}=1,\label{eq: serie geometrica  risolta}
\end{equation}
that is
\begin{equation}
\alpha\equiv\alpha_{m}=\frac{1-X}{1-X^{m}}=\frac{1-\exp\left(-16\pi\right)}{1-\exp\left(-16\pi m\right)}.\label{eq: mio alpha number}
\end{equation}
Hence, one finds that the pre-factor $\alpha$ depends on the BH quantum
level $m.$ Now, if one inserts this result in Eq. (\ref{eq: Corda Probability Intriguing}), one fixes the probability
of emission between the two levels $m$ and $n$ as
\begin{equation}\label{eq: Corda Probability Intriguing finalized}
\Gamma_{m\rightarrow n}=\alpha_{m}\exp\left[16\pi\left(n-m\right)\right]=\left[\frac{1-\exp\left(-16\pi\right)}{1-\exp\left(-16\pi m\right)}\right] \exp\left[16\pi\left(n-m\right)\right].
\end{equation}
From the quantum mechanical point of view, one physically interprets
Hawking radiation like energies of quantum jumps among the unperturbed
levels. One also notice that, for large $m$, it is 
\begin{equation}
\alpha_{m}\simeq \text{constant}=1-\exp\left(-16\pi\right).\label{eq: alfa circa costante}
\end{equation}
Thus, Eq. (\ref{eq: Corda Probability Intriguing finalized}) is well
approximated by 
\begin{equation}\label{eq: probability large m}
\Gamma_{m\rightarrow n}=\alpha_{m}\exp\left[16\pi\left(n-m\right)\right]\simeq\left[1-\exp\left(-16\pi\right)\right]\exp\left[16\pi\left(n-m\right)\right].
\end{equation}
One can also rewrite Eqs. (\ref{eq: 95}) and (\ref{eq: 96}) in terms
of the effective mass
\begin{equation}
\left[M_{E}\right]_{m\rightarrow n}=\frac{M_{m}+M_{n}}{2}\label{eq: massa di transizione}
\end{equation}
which is associated to the transition $m\rightarrow n$, with $n<m$, as
\begin{align}
\braket{n}_{boson}&=\frac{1}{\exp\left[\left(8\pi\left[M_{E}\right]_{m\rightarrow n}\right)\omega\right]-1},\label{eq: final distributions efficace 1}\\
\braket{n}_{fermion}&=\frac{1}{\exp\left[\left(8\pi\left[M_{E}\right]_{m\rightarrow n}\right)\omega\right]+1}.
\label{eq: final distributions efficace 2}
\end{align}

In order to finalize the discussion of this Section, it is important
to raise a very important point. In previous discussion, fixed two
quantum levels $m$ and $n$, it has been considered the emission
of Hawking quanta through the variation of the BH mass-energy 
\begin{equation}
\Delta M_{m\rightarrow n}\equiv M_{n}-M_{m}=2\left(\sqrt{n}-\sqrt{m}\right)=-\omega_{m\rightarrow n}\label{eq: mass variation}
\end{equation}
in an arbitrary transition $m\rightarrow n$, with $n<m$. But in
this paper (and in the Authors' knowledge for the first time in the
literature) it has been shown that the total BH energy $E=-\frac{M_{E}}{2}$
is\emph{ }negative and different from the BH inert mass $M,$ see
Section II for details. Thus, in correspondence of the variation of
the BH inert mass $\Delta M_{m\rightarrow n}$, one must also consider
the variation of the BH total energy, which can be calculated through
Eq. (\ref{eq: BH energy levels finale.}) as 
\begin{equation}
\Delta E_{m\rightarrow n}\equiv-\frac{1}{2}\left(\sqrt{n}-\sqrt{m}\right).\label{eq: energy variation}
\end{equation}
Thus, the total negative energy which is carried out by Hawking radiation
when the BH is subjected to an arbitrary transition $m\rightarrow n$ will be $-\Delta E_{m\rightarrow n}.$ In other words,
the total energy associated to the emissions of Hawking quanta is
negative. Let us clarify this issue. One starts to consider the absorption
of an external particle which generates a quantum transition; in a classical framework, a particle having frequency $\omega$
which falls into a SBH has also a total energy $\omega+E_{g},$ where \cite{key-4}
\begin{equation}
E_{g}\equiv\frac{\omega}{2}\ln\left(1-\frac{2M}{r}\right),\label{eq: energia d'interazione gravitazionale classica}
\end{equation}
is the gravitational energy and $M$ is the BH mass. One sees that
$E_{g}$ diverges to $-\infty$ when the particle approaches the gravitational
radius. From the quantum point of view (see previous Sections), one considers the variation of the BH mass
\begin{equation}
\omega_{m\rightarrow n}=\Delta M_{n\rightarrow m}=-\Delta M_{m\rightarrow n}=2\left(\sqrt{m}-\sqrt{n}\right)\label{eq: opposite mass variation}
\end{equation}
in an arbitrary transition $n\rightarrow m$, with $n<m$. This implies a variation of the BH total energy
\begin{equation}
\Delta E_{n\rightarrow m}\equiv-\Delta E_{m\rightarrow n}=\frac{1}{2}\left(\sqrt{n}-\sqrt{m}\right).\label{eq: opposite energy variation}
\end{equation}
In Eq. (\ref{eq: opposite mass variation}) $\omega_{m\rightarrow n}=\Delta M_{n\rightarrow m}$
is the kinetic energy of the particle which falls into the BH. On
the other hand, such a particle has a total energy given by $\omega_{m\rightarrow n}+E_{g\left(n\rightarrow m\right)}.$
Hence, the quantum version of (\ref{eq: energia d'interazione gravitazionale classica})
is
\begin{equation}
E_{g\left(n\rightarrow m\right)}=\Delta E_{n\rightarrow m}-\omega_{m\rightarrow n}=\frac{5}{2}\left(\sqrt{n}-\sqrt{m}\right).
\label{eq: energia d'interazione gravitazionale}
\end{equation}
This is the gravitational energy of the interaction between the particle
and the BH which, being at the excited state $n,$ will have a mass
$M_{n}$ and a corresponding gravitational radius $2M_{n}$. In fact,
our quantum analysis shows that an absorption of an external particle
having mass $\Delta M_{n\rightarrow m}$ increases the total BH energy
of a negative quantity $\Delta E_{n\rightarrow m}$ given by Eq. (\ref{eq: opposite energy variation}).
Thus, the maximum quantity of the gravitational energy that can be
absorbed by the BH when it absorbs the external particle is given
by Eq. (\ref{eq: energia d'interazione gravitazionale}). A completely
symmetric analysis works in the case of an emission; in that case,
the BH mass decreases of the quantity given by Eq. (\ref{eq: mass variation}),
its total energy decreases of a quantity given by Eq. (\ref{eq: energy variation})
and its gravitational energy decreases of the quantity 
\begin{align}
E_{g\left(m\rightarrow n\right)}&=\Delta E_{m\rightarrow n}+\omega_{m\rightarrow n}\nonumber \\
&=-\frac{5}{2}\left(\sqrt{n}-\sqrt{m}\right),
\label{eq: energia d'interazione gravitazionale emissione}
\end{align}
which means that a negative gravitational energy
is associated to the emitted Hawking quantum having positive kinetic
energy
\begin{equation}
\omega_{m\rightarrow n}=2\left(\sqrt{m}-\sqrt{n}\right).\label{eq: energia positiva}
\end{equation}
In the standard calculations concerning Hawking radiation the negative
gravitational energy is implicitly taken into account by considering
that the emitted Hawking quanta are subjected to the BH strong gravitational
redshift. Thus, when one refers to ``Hawking quanta having positive
energies'', one refers to kinetic energies. 

In summary, the analysis in this paper allows to obtain some very interesting
insights also on the gravitational energy, which represents one of
the more mysterious and controversial issues of gravitational physics
\cite{key-4}. 

\section{Solution to the black hole information paradox}

In 1976 Stephen Hawking claimed that \emph{``Because part of the information
about the state of the system is lost down the hole, the final situation
is represented by a density matrix rather than a pure quantum state}''
(Verbatim from ref. \cite{key-51}). This was the starting point of
the popular ``BH information paradox''. After Hawking's original
claim, enormous time and effort was and is currently devoted to
solve the paradox. Consequences of the BH information paradox are
indeed not trivial. By assuming that information is loss in BH evaporation,
pure quantum states arising from collapsed matter would decay into
mixed states arising from BH evaporation \cite{key-52}. The devastating
consequence is that quantum gravity should not be unitary \cite{key-52}.
Various physicists worked and currently work on this
issue. Some of them remain convinced that quantum information is destroyed
in BH evaporation. Other ones claim that Hawking's original statement
was wrong and information must be, instead, preserved.

In this Section the solution which has been derived in \cite{key-9}
will be refined and adapted to the SBH quantum equations that have
been obtained in previous Sections. 
Let us consider the instant when
the BH Hawking temperature will become higher than the CBR temperature.
At that point, the BH will start to emit Hawking radiation. Thus,
one assumes a first emission from the BH excited state corresponding
to the BH principal quantum number, say $m$, which has a total energy
$E_{m}=-\frac{1}{2}\sqrt{m},$ given by Eq. (\ref{eq: BH energy levels finale.}),
and a corresponding total mass $M_{m}=2\sqrt{m}$,
to a state with $n<m$, which corresponds to a total energy $E_{n}=-\frac{1}{2}\sqrt{n}$
and to a total mass $M_{n}=2\sqrt{n}<M_{m}.$ From the
quantum mechanical point of view, one physically interprets Hawking
radiation like energies of quantum jumps among the unperturbed levels
of Eq. (\ref{eq: BH energy levels finale.}) \cite{key-7,key-9,key-50}.
In quantum mechanics, time evolution of perturbations can be described
by an operator \cite{key-9,key-56}
\begin{equation}\label{eq: perturbazione}
U(t)= \begin{cases} W(t) & \mbox{for } \leq t\leq\tau \\ 0 & \mbox{for } t<0\hspace{0.1cm}\text{and}\hspace{0.1cm}t>\tau\end{cases}   
\end{equation}
Then, the complete (time dependent) Hamiltonian is described by the
operator \cite{key-9,key-56}
\begin{equation}
H(r,t)\equiv V(r)+U(t),\label{eq: Hamiltoniana completa}
\end{equation}
where $V(r)$ is given by Eq. (\ref{eq: energia potenziale BH effettiva}).
Thus, considering a wave function $\psi(r,t),$ one can write the
correspondent time dependent Schroedinger equation for the
system:
\begin{equation}\label{eq: Schroedinger equation}
i\frac{d\ket{\psi(r,t)}}{dt}=\left[V(r)+U(t)\right]\ket{\psi(r,t)}=H(r,t)\ket{\psi(r,t)}.
\end{equation}
The\emph{ }state\emph{ }which satisfies Eq. (\ref{eq: Schroedinger equation})
is
\begin{equation}
\ket{\psi(r,t)}=\sum_{n}a_{n}(t)\exp\left(-iE_{n}t\right)\ket{\varphi_{n}(r)},\label{eq: Schroedinger wave-function}
\end{equation}
where the $\varphi_{n}(r)$'s are the eigenfunctions of the time independent
Schr\"odinger equation (\ref{eq: Schrodinger equation BH effettiva})
and the $E_{n}$ are the correspondent eigenvalues. In the basis
$\ket{\varphi_{n}(r)}$, the matrix elements of $W(t)$ can be written
as 
\begin{equation}
W_{ij}(t)\equiv A_{ij}\delta(t),\label{eq: a delta}
\end{equation}
where $W_{ij}(t)=\braket{\varphi_{i}(r)|W(t)|\varphi_{j}(r)}$ and the $A_{ij}$'s
are real. In order to solve the complete quantum mechanical problem
described by the operator (\ref{eq: Hamiltoniana completa}), one
needs to know the probability amplitudes $a_{n}(t)$ due to the application
of the perturbation described by the time dependent operator (\ref{eq: perturbazione}),
which represents the perturbation associated to the emission of a
Hawking quantum. For $t<0,$ i.e., before the perturbation operator
(\ref{eq: perturbazione}) starts to work, the system is in a stationary
state $\ket{\varphi_{m}(t,r)},$ at the quantum level $m$, with energy
$E_{m}=-\frac{1}{2}\sqrt{m},$.
Thus, in Eq. (\ref{eq: Schroedinger wave-function}) only the term 
\begin{equation}
\ket{\psi_{m}(r,t)}=\exp\left(-iE_{m}t\right)|\varphi_{m}(r)>,\label{eq: Schroedinger wave-function in.}
\end{equation}
is non-zero for $t<0$. This implies $a_{n}(t)=\delta_{nm}\:\:$for
$\:t<0.$ When the perturbation operator (\ref{eq: perturbazione})
stops to work, i.e., after the emission, for $t>\tau$ the probability
amplitudes $a_{n}(t)$ return to be time independent, having the value
$a_{m\rightarrow n}(\tau)$. In other words, for $t>\tau\:$ the system
is described by the wave function $\psi_{f}(r,t)$, which
corresponds to the state 
\begin{equation}
\ket{\psi_{f}(r,t)}=\sum_{n=1}^{m}a_{m\rightarrow n}(\tau)\exp\left(-iE_{n}t\right)\ket{\varphi_{n}(x)}.\label{eq: Schroedinger wave-function fin.}
\end{equation}
Therefore, the probability to find the system in an eigenstate having
energy $E_{n}=-\frac{1}{2}\sqrt{n}$, with $n<m$ for emissions, is given by 
\begin{equation}
\Gamma_{m\rightarrow n}(\tau)=|a_{m\rightarrow n}(\tau)|^{2}.\label{eq: ampiezza e probability}
\end{equation}
By using a standard analysis, one obtains the following differential
equation from Eq. (\ref{eq: Schroedinger wave-function fin.}):
\begin{equation}
i\dot{a}_{m\rightarrow n}(t)=\sum_{l=1}^{n}W_{ml}a_{m\rightarrow l}(t)\exp\left[i\left(\Delta E_{l\rightarrow n}\right)t\right],\label{eq: systema differenziale}
\end{equation}
where the dot over $a_{m\rightarrow n}(t)$ denotes the derivative with respect to time.
To first order in $U(t)$, by using the Dyson series \cite{key-56},
one gets the following solution:
\begin{equation}
a_{m\rightarrow n}=-i\int_{0}^{t}\left\{ W_{nm}(t')\exp\left[i\left(\Delta E_{m\rightarrow n}\right)t'\right]\right\} dt'.\label{eq: solution}
\end{equation}
By inserting Eq. (\ref{eq: a delta}) in Eq. (\ref{eq: solution})
one obtains 
\begin{align}\label{eq: solution 2}
a_{m\rightarrow n}=iA_{nm}\int_{0}^{t}\left\{ \delta(t')\exp\left[i\left(\Delta E_{m\rightarrow n}\right)t'\right]\right\} dt'=\frac{i}{2}A_{nm}.
\end{align}
Combining this equation with Eqs. (\ref{eq: Corda Probability Intriguing})
and (\ref{eq: ampiezza e probability}) one obtains
\begin{equation}\label{eq: uguale}
A_{nm}=2\sqrt{\alpha}\exp\left[8\pi\left(n-m\right)\right],
\end{equation}
and so
\begin{equation}
a_{m\rightarrow n}=-i\sqrt{\alpha}\exp\left[8\pi\left(n-m\right)\right].
\end{equation}
As $\sqrt{\alpha}\sim1,$ one gets $A_{nm}\sim10^{-11}$ for $n=m-1$,
i.e., when the probability of emission has its maximum value. This
implies that second order terms in $U(t)$ can be neglected. Clearly, for $n<m-1$, the approximation is better
because the $A_{mn}$'s are even smaller than $10^{-11}$. Thus, one
can write down the final form of the ket representing the state as
\begin{equation}
\ket{\psi_{f}(r,t)}=\sum_{n=1}^{m}a_{m\rightarrow n}\exp\left[-iE_{n}t\right]\ket{\varphi_{n}(r)}.\label{eq: Schroedinger wave-function finalissima}
\end{equation}
Eq. (\ref{eq: Schroedinger wave-function finalissima})
represents a \emph{pure final state instead of a mixed final state.}
Then, the states are written in terms of a \emph{unitary} evolution
matrix instead of a density matrix and this implies
the fundamental conclusion that information is not loss in
BH evaporation. This result is consistent with 't Hooft's idea that
Schr\"odinger equations can be used universally for all dynamics in
the universe \cite{key-57} and dismisses the claim of Hawking \cite{key-51}
that has been cited at the starting of this Section. The above final state is due to potential arbitrary transitions $m\rightarrow n$, with $m>n$.
Then, the subsequent \emph{collapse of the wave function} to a new
stationary state, at the quantum level $n$,
\begin{equation}
\ket{\psi_{n}(r,t)}=\exp\left(-iE_{n}t\right)\ket{\varphi_{n}(r)},\label{eq: Schroedinger wave-function out}
\end{equation}
implies that the wave function of the infalling particle in Hawking's
mechanism of particles creation by BH has been transferred to the
two-particle system described in the previous Sections, and it is given by 
\begin{equation}\label{eq: funzione onda particella emessa}
\ket{\psi_{\left(m\rightarrow n\right)}(r,t)}\equiv\exp\left(-iE_{n}t\right)\ket{\varphi_{n}(r)}-\exp\left(-iE_{m}t\right)\ket{\varphi_{m}(r)}.
\end{equation}
This wave function results entangled with the wave function of the
particle which propagates towards infinity. Now, it will be shown
that this key point solves the entanglement problem connected with
the information paradox. This problem
concerns the entanglement structure of the wave function associated
to the particle pair creation \cite{key-54}. In other terms, in order
to solve the paradox, one needs to know the part of the wave function
in the interior of the horizon, i.e., the part of the
wave function associated to the interior, infalling mode. This is
exactly the part of the wave function which in the Hawking computation
gets entangled with the part of the wave function outside, i.e., the
part of the wave function associated to the particle which escapes
from the BH \cite{key-54}. Here the key point is that
the particle which falls into the BH transfers its part of the wave
function and, in turn, the information encoded in it, to the two-particle system governed by Eqs. (\ref{eq: energia potenziale BH effettiva}) - (\ref{eq: energia totale effettiva}). Hence, the emitted radiation results to be entangled with such a quantum system which concerns the ``excited
states'' of the ``gravitational atom''. Thus, one argues that the
BH response to the absorption of an interior, infalling mode is to
add energy to the ``atom's excited state'' corresponding to
the energy level $E_{m}$, in order to allow it to jump to the ``atom's
excited state'' corresponding to the energy level $E_{n}$. In that
way, the interior part of the wave function is now ``within'' the
two-particle system, which is now at the
quantum level $E_{n}$. Let us clarify this point. Again, let
us consider the instant when the BH Hawking temperature is
higher than the CBR temperature. The BH will be in a stationary state
given by Eq. (\ref{eq: Schroedinger wave-function in.}). Then, the
BH will start to emit Hawking radiation and one assumes a first emission
from the BH excited state corresponding to the BH principal quantum
number $n=m$, which has a total energy $E_{m}=-\frac{1}{2}\sqrt{m}$, and a corresponding total
mass $M_{m}=2\sqrt{m},$ given by Eq. (\ref{eq: spettro massa BH finale}),
to a state with $n=m_{1}<m$, which corresponds to a total energy
$E_{m_{1}}=-\frac{1}{2}\sqrt{m_{1}}$ and to a total mass $M_{m_{1}}=2\sqrt{m_{1}}.$
It will be $M_{m_{1}}<M_{m}$ because the infalling particle in Hawking's
mechanism of particles creation by BH has negative mass. The energy
jump between the two levels is 
\begin{equation}
\Delta E_{m\rightarrow m_{1}}\equiv E_{m_{1}}-E_{m}=\frac{1}{2}\left(\sqrt{m}-\sqrt{m_{1}}\right)\label{eq: salto}
\end{equation}
In other words, the energy of the first absorbed particle having negative
mass is transferred, together with its part of the wave function,
to the two-particle system, which is now entangled
with the emitted particle. Now, by using Eq. (\ref{eq: funzione onda particella emessa}),
and by setting $n=m_{1}$, one finds that the part of the wave function
in the interior of the horizon is
\begin{equation}\label{eq: da zero ad n1}
\ket{\psi_{\left(m\rightarrow m_{1}\right)}(r,t)}\equiv\exp\left(-iE_{m_{1}}t\right)\ket{\varphi_{m_{1}}(r)}-\exp\left(-iE_{m}t\right)\ket{\varphi_{m}(r)}.
\end{equation}
Let us consider a second emission, which corresponds to the transition
from the state with $n=m_{1}$ to a state with, say, $n=m_{2}<m_{1}$.
The BH total energy changes from $E_{m_{1}}=-\frac{1}{2}\sqrt{m_{1}}$ to 
$E_{m_{2}}=-\frac{1}{2}\sqrt{m_{2}}$, while the BH mass changes from $M_{m_{1}}=2\sqrt{m_{1}}$ to $M_{m_{2}}=2\sqrt{m_{2}}<M_{m_{1}}.$
The energy jump between the two levels is
\begin{equation}
\Delta E_{m_{1}\rightarrow m_{2}}\equiv E_{m_{2}}-E_{m_{1}}=\frac{1}{2}\left(\sqrt{m_{1}}-\sqrt{m_{2}}\right).\label{eq: salto 2}
\end{equation}
As before, the energy of the second absorbed particle having negative mass is
transferred, together with its part of the wave function, to
the two-particle system, which is now entangled
with both of the emitted particles. By using again Eq. (\ref{eq: funzione onda particella emessa})
and setting $n=m_{2}$ and $m=m_{1}$, one finds that the part of
the wave function of the second infalling mode is
\begin{equation}\label{eq: da n1 ad n2}
\ket{\psi_{\left(m_{1}\rightarrow m_{2}\right)}(r,t)}\equiv\exp\left(-iE_{m_{2}}t\right)\ket{\varphi_{m_{2}}(r)}-\exp\left(-iE_{m_{1}}t\right)\ket{\varphi_{m_{1}}(r)}.
\end{equation}
Let us consider a third emission, which corresponds to the transition
from the state with $n=m_{2}$ to a state with, say, $n=m_{3}<m_{2}$.
The BH total energy changes from $E_{m_{2}}=-\frac{1}{2}\sqrt{m_{2}}$ to 
$E_{m_{3}}=-\frac{1}{2}\sqrt{m_{3}}$, while the BH mass changes from $M_{m_{2}}=2\sqrt{m_{2}}$ to $M_{m_{3}}=2\sqrt{m_{3}}<M_{m_{2}}.$
The energy jump between the two levels is
\begin{equation}
\Delta E_{m_{2}\rightarrow m_{3}}\equiv E_{m_{3}}-E_{m_{2}}=\frac{1}{2}\left(\sqrt{m_{2}}-\sqrt{m_{3}}\right).\label{eq: salto 3}
\end{equation}
The energy of the third absorbed particle having negative mass is
transferred to the two-particle, which is now entangled
with the third emitted particles. By using again Eq. (\ref{eq: funzione onda particella emessa}),
and setting $n=m_{3}$ and $m=m_{2}$, one finds that the part of
the wave function of the third infalling mode is
\begin{equation}\label{eq: da n2 ad n3}
\ket{\psi_{\left(m_{2}\rightarrow m_{3}\right)}(r,t)}\equiv\exp\left(-iE_{m_{3}}t\right)\ket{\varphi_{m_{3}}(r)}-\exp\left(-iE_{m_{2}}t\right)\ket{\varphi_{m_{2}}(r)}.
\end{equation}
The process will continue again and again till the
\emph{Planck distance} and the \emph{Planck mass} are approached by
the evaporating BH. At that point, as it has been stressed in previous
Sections, the GUP prevents the total BH evaporation in exactly the
same way that the HUP prevents the hydrogen atom from total collapse, and the two-particle system
is entangled with all the Hawking quanta emitted at that time. Hence,
the BH arrives at the minimum energy level which is compatible with the
GUP. Such an energy level corresponds to Eq. (\ref{eq: energia minima})
and consists of a negative energy having the absolute value of the
Planck energy. In other words, subsequent ``absorptions'' by the
BH of the wave functions of absorbed particles having negative mass-energy
in Hawking's mechanism lead the total BH wave
function, which at the starting of the emission process was given by
Eq. (\ref{eq: Schroedinger wave-function in.}), to evolve till the BH
wave-function of the minimum energy level is 
\begin{equation}
\ket{\psi_{1}(r,t)}=\exp\left(-iE_{1}t\right)\ket{\varphi_{1}(r)}.\label{eq: wave-function minimum}
\end{equation}
Clearly, the evolution of BH evaporation that it has been discussed
above is \emph{unitary}.

Therefore, it has been shown that SBHs, which are considered the fundamental
bricks of quantum gravity, are well defined quantum mechanical systems,
having ordered, discrete quantum spectra, which preserve physical
information by restoring predictability in gravitational collapse. 

\section{Conclusion remarks}
Rosen's approach has been originally applied to the historical Oppenheimer
and Snyder gravitational collapse in \cite{key-5}. By setting the constraints for the formation of the SBH, the gravitational potential, the Schr\"odinger equation, the solution for the energy levels,
and the area quantum have been found, by also discussing the quantum representation of the BH's ground state at the Planck scale. Such results are
consistent with previous ones in the literature. It was also shown
that the traditional classical singularity in the core of the SBH
is replaced by a non-singular two-particle system where the two components,
the ``nucleus'' and the ``electron'', strongly interact with each
other.  In agreement with the de Broglie hypothesis \cite{de Broglie}, the ``electron'' has been interpreted in terms of the quantum oscillations of the BH horizon. In Section
II, the results obtained in \cite{key-5} have been reviewed by also adding some new insights; in Section III it has been indeed shown that the same results can be obtained through a path
integral approach \cite{key-23}; in Section IV it has been shown
that such results allow to compute the SBH entropy as a function
of the BH principal quantum number in terms of Bekenstein-Hawking
entropy and three sub-leading corrections. In addition, the coefficient of the formula of
Bekenstein-Hawking entropy is reduced to a quarter of its
traditional value; in Section V it has been shown that, by
performing a correct rescaling of the energy levels, the semi-classical
Bohr-like approach to BH quantum physics, previously developed by
one of the Authors (CC) in \cite{key-8,key-9,key-21,key-24,key-25},
is consistent with the obtained results for large values of the BH
principal quantum number. After this, Hawking radiation has been analysed
by discussing its connection with the BH quantum structure in Section
VI; finally, in Section VII, by analyzing
the time evolution of the GHA, the solution of the BH information paradox is discussed. In fact, on one hand, it has been shown
that, contrary to the famous claim of Hawking in \cite{key-51}, the
final situation of BH evaporation is represented by a pure quantum
state given by Eq. (\ref{eq: Schroedinger wave-function finalissima}).
On the other hand, also the entanglement problem connected with the
information paradox has been solved because emitted Hawking radiation
results entangled with the two-particle system governed by Eqs. (\ref{eq: energia potenziale BH effettiva}) - (\ref{eq: energia totale effettiva})
in a unitary process which allows the BH total wave function at the
starting of the emission process to evolve till the BH wave-function
of the minimum energy level.
Therefore, the results obtained here seem consistent with the interesting approach of Hajicek and Kiefer (see \cite{key-68,key-69} for details). In such works, the Authors indeed discussed the quantization of a spherical dust shell by constructing a well-defined self-adjoint extension for the Hamilton operator. As a result,
the evolution is unitary and the singularity is avoided.

Finally one takes the chance to recall that, in a series of interesting papers \cite{key-62,key-63,key-64}, the Authors wrote down the Schr\"odinger equation for a collapsing object and showed by explicit calculations that quantum mechanics is perhaps able to remove the singularity at the BH center (in various space-time slicings); this is consistent with our analysis. Moreover, they also proved (among the other things) that the wave function of the collapsing object is non-singular at the center even when the radius of the collapsing object (classically) reaches zero. In \cite{key-64}, they considered charged BHs. 

In regard to the area quantization, another interesting approach, based on graph theory, can be found in \cite{key-65}. Here, the Bekenstein-Hawking area entropy accompanied with a proper logarithmic term (subleading correction) is obtained, and the size of the unit horizon area is fixed. 

Curiously, Davidson also found a hydrogen-like spectrum in a totally different contest \cite{key-66}. 
It seems that the results found in this paper are in agreement with the previous literature.

\section*{Acknowledgments}
One of the Authors (F.T.) gratefully acknowledges ZKM and Peter Weibel for the financial support.

\end{document}